\documentclass[12pt, a4paper]{article}

\usepackage[height=22cm,width=16cm, centering]{geometry}
\usepackage{setspace}

\usepackage[utf8]{inputenc}

\usepackage{amsmath}
\usepackage{amssymb}
\usepackage[dvipdfmx]{graphicx}
\usepackage{subcaption}
\usepackage{color,comment, bm}

\usepackage[sort&compress, numbers, merge]{natbib}

\definecolor{purple}{rgb}{0.5 ,0, 0.7}
\definecolor{bluegreen}{rgb}{0, 0.45, 0.35}
\usepackage{hyperref}
\hypersetup{colorlinks=true, linkcolor=bluegreen, citecolor=purple, urlcolor=blue}

\definecolor{DarkBlue}{rgb}{0,0,0.9} 

\definecolor{DarkRed}{rgb}{0.65,0,0}

\allowdisplaybreaks

\begin{document}

\begin{titlepage}
\begin{center}
\leavevmode \\
\vspace{ -3cm}

\hfill{\scriptsize RUP-22-25}\\ \vspace{-2mm}
 \hfill{\scriptsize KEK-TH-2365}\\ \vspace{-2mm}
 \hfill{\scriptsize KEK-Cosmo-0280}\\ \vspace{-2mm}
 \hfill{\scriptsize KEK-QUP-2022-0011}\\ \vspace{-2mm}
 \hfill{\scriptsize YITP-22-143}\\ \vspace{-2mm}
\hfill {\scriptsize CTPU-PTC-22-27}\\

\noindent

\vglue .05in
  {\Large Threshold of Primordial Black Hole Formation against \\ \vspace{0.5mm} Velocity Dispersion in Matter-Dominated Era}
  
\vglue .15in

{
Tomohiro Harada$^{a}$, Kazunori Kohri$^{b, c, d, e}$, Misao Sasaki$^{e, f, g}$, \\ Takahiro Terada$^{h}$, and Chul-Moon Yoo$^{i}$
}

\vglue.15in

\scriptsize
\textit{ 
${}^{a}$ \,  Department of Physics, Rikkyo University, Toshima, Tokyo 171-8501, Japan \\
${}^{b}$ \,  Institute of Particle and Nuclear Studies, KEK,
1-1 Oho, Tsukuba, Ibaraki 305-0801, Japan \\
${}^{c}$ \,  International Center for Quantum-field Measurement Systems for Studies of the Universe
and Particles (QUP, WPI), KEK, 1-1 Oho, Tsukuba, Ibaraki 305-0801, Japan \\
${}^{d}$ \,The Graduate University for Advanced Studies (SOKENDAI), \\
1-1 Oho, Tsukuba, Ibaraki 305-0801, Japan\\
${}^{e}$ \,Kavli Institute for the Physics and Mathematics of the Universe (WPI),\\
University of Tokyo, Kashiwa 277-8583, Japan \\
${}^{f}$ \,Center for Gravitational Physics, Yukawa Institute for Theoretical Physics, \\ Kyoto University, Kyoto 606-8502, Japan \\
${}^{g}$ \,Leung Center for Cosmology and Particle Astrophysics, \\
National Taiwan University, Taipei 10617, Taiwan \\
${}^{h}$ \,Center for Theoretical Physics of the Universe,\\ Institute for Basic Science (IBS), Daejeon, 34126, Korea \\
 ${}^{i}$ \,Division of Particle and Astrophysical Science, Graduate School of Science, \\ Nagoya University, Nagoya 464-8602, Japan \\
}
\normalsize

\end{center}

 \vglue 0.15in

\begin{abstract}
We study the effects of velocity dispersion on the formation of primordial black holes~(PBHs) in a matter-dominated era. 
The velocity dispersion is generated through the nonlinear growth of perturbations
and has the potential to impede the gravitational collapse and thereby the formation of PBHs.
To make discussions clear, we consider two distinct length scales. The larger one is where gravitational collapse occurs which could lead to PBH formation, 
and the smaller one is where the velocity dispersion develops due to nonlinear interactions.
We estimate the effect of the velocity dispersion on the PBH formation
by comparing the free-fall timescale and the timescale for a particle to cross the collapsing region.
As a demonstration, we consider a log-normal power spectrum for the initial density perturbation with the peak value $\sigma_0^2$ at a scale
that corresponds to the larger scale.
We find that the threshold value of the density perturbation $\tilde \delta_{\rm th}$ at the horizon entry for the PBH formation scales as
$\tilde \delta_{\rm th}\propto \sigma_0^{2/5}$ for $\sigma_0\ll1$.
\end{abstract}

\end{titlepage}

\newpage


\section{Introduction}

Black holes are widely accepted to exist in the Universe through various observations such as  
the motion of surrounding celestial objects~\cite{Ghez:2008ms, Gillessen:2009ht, Genzel:2010zy, GRAVITY:2018ofz}, 
gravitational-wave signals from binary-black-hole mergers~\cite{LIGOScientific:2016aoc, LIGOScientific:2018mvr, LIGOScientific:2021usb}, 
and the black hole shadow~\cite{EventHorizonTelescope:2019dse}. 
Some of the black holes might have a primordial origin~\cite{Zeldovich1966, Hawking:1971ei, Carr:1974nx}.  
Depending on their mass, primordial black holes (PBHs) may explain dark matter abundance~\cite{Carr:2009jm,Carr:2020gox,Carr:2020xqk,Escriva:2022duf}, 
the binary black hole abundance observed by the gravitational-wave events~\cite{Bird:2016dcv, Clesse:2016vqa, Sasaki:2016jop,Sasaki:2018dmp}, 
an excess of microlensing events~\cite{Niikura:2019kqi}, 
a potential gravitational-wave signal at 
a pulsar timing array~\cite{NANOGrav:2020bcs, Vaskonen:2020lbd, DeLuca:2020agl, Kohri:2020qqd, Sugiyama:2020roc, Domenech:2020ers, Inomata:2020xad}, 
the seeds of supermassive black holes and cosmic structures~\cite{Kawasaki:2012kn,Kohri:2014lza,Nakama:2016kfq,Carr:2018rid,Serpico:2020ehh,Unal:2020mts,Kohri:2022wzp}.  
PBHs whose lifetime is shorter than the age of the Universe may also play important roles 
in the early Universe such as baryogenesis~\cite{Toussaint:1978br, Barrow:1990he, Baumann:2007yr} 
and the production of stochastic gravitational-wave 
background~\cite{Inomata:2020lmk, Papanikolaou:2020qtd, Domenech:2020ssp, Domenech:2021wkk, Bhaumik:2020dor, Hooper:2020evu, Inomata:2019ivs, Papanikolaou:2022chm}.\footnote{
The stochastic background of the gravitational waves induced by the primordial curvature perturbations is also associated with PBHs 
with longer lifetimes and they are interesting observational targets.  
See Refs.~\cite{Ananda:2006af, Baumann:2007zm, Saito:2008jc, Saito:2009jt, Assadullahi:2009nf, Espinosa:2018eve, Kohri:2018awv, Cai:2018dig} 
and recent reviews~\cite{Yuan:2021qgz, Domenech:2021ztg}.
}

For any applications of PBHs, their abundance is a basic quantity that determines their cosmological significance. 
For the estimation of the abundance in a radiation-dominated (RD) era, three tools conventionally adopted are: the criterion of PBH formation~%
\cite{Shibata:1999zs, Musco:2004ak,  Polnarev:2006aa, Musco:2008hv, 
Musco:2012au, Nakama:2013ica, Harada:2013epa, Harada:2015yda, Musco:2018rwt, Escriva:2019phb, Musco:2020jjb, Musco:2021sva, Papanikolaou:2022cvo},
the peaks theory for the statistical treatment of the initial profiles%
~\cite{1986ApJ...304...15B,Yoo:2018kvb,Germani:2018jgr,Yoo:2020dkz}, and 
the critical behavior of the PBH formation~\cite{Niemeyer:1997mt, Yokoyama:1998xd, Green:1999xm, Kuhnel:2015vtw} for sufficiently spherical collapses.  
The latest theoretical estimation of the PBH abundance for the monochromatic curvature power spectrum 
can be seen in Ref.~\cite{Kitajima:2021fpq} for Gaussian and local-type non-Gaussian curvature perturbations (see also Refs.~\cite{Atal:2019erb,Yoo:2019pma,Atal:2019cdz,Kitajima:2021fpq,Escriva:2022pnz} for effects of the non-Gaussianity), 
where the PBH mass function is calculated based on 
an approximately universal criterion~\cite{Escriva:2019phb} and the peaks theory~\cite{Yoo:2020dkz}.

In a matter-dominated (MD) era, since the pressure of the fluid, 
which is the most effective factor to impede PBH formation in an RD era, is absent and 
the criterion of PBH formation obtained with the fluid approximation does not apply. 
The absence of the pressure would make the gravitational collapse easier, 
and lower peaks of the density perturbation may contribute to PBH abundance~\cite{Khlopov:1980mg, Polnarev:1986}. 
However, the high-peak approximation in peaks theory becomes worse and the spherical symmetry assumption, 
which is relevant to rare high peaks~\cite{1986ApJ...304...15B}, may not be accurate. 
Furthermore, it is well known that the deviation from the spherical symmetry can significantly grow 
in the gravitational collapse in the case of pressureless matter~\cite{1965ApJ...142.1431L,1970A&A.....5...84Z}. 
Thus, non-spherical initial profiles are inevitable, and they may lead to the complicated nonlinear dynamics 
associated with deformation and rotation. In addition, the pressureless fluid approximation breaks down as soon as
nonlinearity becomes important due to the occurrence of shell-crossing singularities.
In reality, the pressureless matter obeys the collisionless Boltzmann equation, and velocity dispersion will develop
in the nonlinear regime.
Thus, careful studies that take account of various effects are required in order to understand the PBH formation in an MD era.  
There are several analytical studies in the literature, 
in which the effects of asphericity~\cite{Harada:2016mhb}, angular momentum~\cite{Harada:2017fjm}, 
and inhomogeneity~\cite{Khlopov:1980mg, Polnarev:1986, Kokubu:2018fxy}, were estimated 
under some assumptions and approximations. 
For an analysis in numerical relativity, see Ref.~\cite{deJong:2021bbo}.

In this paper, we consider the effect of the velocity dispersion inside the collapsing region and discuss 
the conditions for the velocity dispersion to impede the PBH formation in the MD era.\footnote{
The effect of the velocity dispersion was briefly mentioned in Ref.~\cite{Padilla:2021zgm},
 in which the radius of the virialized structure was compared to the Schwarzschild radius.  We introduce a more comprehensive discussion.
}  
To make discussions simple and clear, we consider two distinct length scales.
The larger one is where gravitational collapse occurs which could lead to PBH formation,  
and the smaller one is where the velocity dispersion develops due to nonlinear interactions.
The paper is organized as follows. 
In Sec.~\ref{sec:overview}, we overview the situation we consider by describing our setup and assumptions.
In Sec.~\ref{sec:collapsecond}, we derive the conditions for the velocity dispersion to impede the PBH formation.
In doing so, we find there are essentially two different scenarios, depending on when the velocity dispersion emerges,
details of which are discussed in Appendix~\ref{sec:generation}.
In Sec.~\ref{sec:threshold}, the conditions to impede the PBH formation are rephrased 
in terms of the density threshold for the PBH formation for a given density power spectrum.
In Sec.~\ref{sec:PBH_abundance}, using the obtained results, 
we compute the PBH production probability as a function of the peak amplitude of the power spectrum of the density perturbation. 
Our conclusions are given in Sec.~\ref{sec:conclusion}.
Throughout this paper, we use the units in which both the speed of light and Newton's gravitational constant are unity, $G=c=1$.

\section{Overview}
\label{sec:overview}

To avoid any confusion in understanding how the velocity dispersion 
is generated and works as a preventing factor against PBH formation, 
before discussing the details, let us clarify our setup and assumptions. 

First, the rough basic picture is as follows.  We consider a
high-density region of the comoving scale $k_\mathrm{PBH}$
that undergoes gravitationally collapse and may lead to PBH formation, which may be
impeded by the velocity dispersion generated by the dynamics on a scale $k$ where
$k>k_\mathrm{PBH}$.  
Here we suppose that a finite power of perturbation also exists on these small scales. 
When the perturbations on the small scales enter the nonlinear regime, that is, when the density perturbation becomes high enough to render the fluid approximation invalid, the velocity dispersion is assumed to be generated, so that the random motion of the constituent particles can act against gravity. 
In this phase, if the repulsive effect due to this generated random motion dominates the dynamics of the whole system over the length scale $1/k_\text{PBH}$, the collapsing region will be eventually virialized, so that the PBH formation will be impeded. Here, we note that 
the velocity dispersion may get effective on the scale $k_\mathrm{PBH}$ not immediately after it is generated on the scale $k$ but later, depending on its magnitude. We describe more specific scenarios in the following setup and assumptions. 

Now let us describe the setup and assumptions. 
\begin{itemize}
\item  
We consider only the two separated comoving scales characterized by $k$ and $k_{\rm PBH}$. 
We call $k_{\rm PBH}$ the large or PBH scale, while $k$ the small scale.
Throughout this paper, 
the tilded quantities are those for the PBH scale $k_{\rm PBH}$, 
while those for the small scale $k$ are not tilded. 
Thus in particular $\tilde k\equiv k_{\rm PBH}$.
Two exceptions are the velocity dispersion $\sigma_v(t)$ as a function of time,
and the root-mean-square amplitude of the density perturbation at the horizon entry $\sigma_{\delta,{\rm ent}}(k)$ as a function of the comoving wavenumber.

\item Since we discuss a cosmological patch that is going to form a PBH (or to be impeded by the velocity dispersion), 
the density perturbation on this patch $\tilde \delta_{\rm ent}$ is exceptionally higher than the typical amplitude,
$\tilde \delta_{\rm ent}\gg\sigma_{\delta,{\rm ent}} (\tilde{k})$, at the time the scale enters the Hubble horizon. 

\item 
On the other hand, we suppose the amplitude of the smaller-scale perturbations is given by a typical value, that is, the amplitude at the horizon entry 
$\delta_{\rm ent}(k)$ is typically given by the standard deviation of the density perturbation, $|\delta_{\rm ent}(k)|=\mathcal O(\sigma_{\delta,{\rm ent}}(k))$.
This magnitude is also assumed to be so small that the PBH formation \emph{on the small scale} is negligible.  
Then, except for Sec.~\ref{sec:PBH_abundance}, we assume for simplicity that the only impeding factor against PBH formation 
on the scale $\tilde{k}$ is the velocity dispersion.  
In addition to the velocity dispersion, other potential impeding factors against the PBH formation will be briefly discussed in Sec.~\ref{sec:PBH_abundance}. 

\item We assume that the fluid approximation (i.e., coherent flow of particles with negligible velocity dispersion) is valid on the scale $k$
until caustics appear, that is, until the perturbation enters the nonlinear regime (provided that larger scales have not yet collapsed).  
  The collapsing overdense regions may be locally described by the closed Friedmann-Lema\^itre-Robertson-Walker (FLRW) universe models (see, e.g., Ref.~\cite{Yoo:2022mzl}).  
Then the collapsing time on the PBH scale $\tilde{k}$ and 
that on the scale $k$ may be estimated by using the closed FLRW
collapsing model, which we denote by 
 $\tilde t_\mathrm{coll}$ and $t_\mathrm{coll}$, respectively. 
It should be noted that, if $t_\mathrm{coll}\gtrsim \tilde t_\mathrm{coll}$, the smaller-scale perturbations 
cannot be independent of the surrounding over-density associated with the PBH scale. 
Namely, the actual collapsing time will be earlier than $t_{\rm coll}$ in that case. 
\end{itemize}

It should be noted that the velocity dispersion generated on the scale $k$ may not be immediately shared with the whole region on 
the scale $\tilde{k}$. If the virialization is completed on the smaller scale before the collapse of the larger scale, 
the matter is confined in the virialized halos. 
In this case, the halos behave as \emph{macroscopic self-gravitating comoving particles}, hence no impeding effect due to the velocity dispersion
is expected against the gravitational collapse on the scale $k_\mathrm{PBH}$. 
Therefore, we consider the two possible scenarios: Case I and Case II as follows. 

\begin{itemize}
 \item Case I: We assume the virialization of the perturbations on the smaller scale is completed before the collapse of the larger scale,
 i.e., $t_\mathrm{vir} < \tilde t_\mathrm{coll}$. In this case, the velocity dispersion is
  released when the mean density on the scale $\tilde{k}$ becomes comparable to the density of a halo.\footnote{
Note that the mean density first decreases as the Universe expands, but eventually increases as the region on the scale $\tilde{k}$ contracts.  
The release of the velocity dispersion happens while the mean density is increasing.}  
At this point, the halos \emph{dissolve} into ambient space. 
We identify the time when the halos dissolve with the time  ($\equiv t_*$) at which the velocity dispersion becomes effective on the PBH scale $\tilde{k}$.

\item Case II: We assume  $t_{\rm vir}>\tilde t_{\rm coll}$, i.e., there is no time for the small scale to collapse and form virialized halos.
In this case, the velocity dispersion is directly released soon after the smaller-scale perturbations grow into the nonlinear regime. 
Since the evolution of the small-scale perturbations should be necessarily promoted (especially in case II) by the evolution 
of the surrounding high-density region on the scale $1/\tilde{k}$, 
we will consider the linear density perturbation $\delta_{\rm com}$ in a FLRW universe in the synchronous comoving gauge for the evolution of the small-scale perturbations. 
Then, the time $t_*$ when the velocity dispersion becomes effective on the scale $\tilde{k}$ is 
identified with the time when $\delta_{\rm com}\sim1$, i.e., when caustics appear. 
\end{itemize}

The detailed derivation of the criterion that distinguishes the two cases is given in Appendix. Here let us briefly present an order-of-magnitude estimate.
Consider a region where a positive density perturbation $\tilde \delta$ exists. In a linear regime, it grows in proportion to the scale factor $a$ at a matter-dominated stage, 
where $a\propto t^{2/3}$. 
Thus $\tilde \delta(t)=(t/\tilde t_{\rm ent})^{2/3}\tilde \delta_{\rm ent}$, where $\tilde \delta_{\rm ent}$ is the amplitude of the density perturbation at horizon entry. 
The time at which the region collapses is sometime after the amplitude reached $\mathcal O(1)$. 
Ignoring order-of-unity numerical factors, we may identify the time when $\tilde \delta=1$ as the collapse time, $\tilde t_{\rm col}$.
When the region collapses but doesn't lead to the formation of a black hole, the system virializes approximately at the same time, $\tilde t_{\rm vir}\sim \tilde t_{\rm col}$.
Since the same argument applies to the small scale, the region of the small scale will be virialized, as $\delta_{\rm ent}$ is assumed to be very small.
Thus we have
 \begin{eqnarray}
 {t}_{\rm vir} \sim {t}_{\rm ent} {\delta}_{\rm ent}^{-3/2}\,.
 \label{t_vir}
 \end{eqnarray}
 On the other hand, on the PBH scale $\tilde{k}$, the region may form a black hole if there is no impeding factors. So we estimate the collapse time as
 \begin{eqnarray}
 	\tilde {t}_{\rm col}\sim \tilde {t}_{\rm ent}\tilde \delta_{\rm ent}^{-3/2}\,.
 \label{t_col}
 \end{eqnarray}
 Now recalling that the comoving scale $k$ may be expressed in terms of its horizon entry time as $k=(aH)_{\rm ent}\propto t_{\rm ent}^{-1/3}$, one finds
 \begin{eqnarray}
\left( \frac{{t}_{\rm vir}}{\tilde t_{\rm col}}\right)^{2/3}
\sim\left(\frac{{t}_{\rm ent}}{\tilde t_{\rm ent}}\right)^{2/3} \frac{\tilde \delta_{\rm ent}}{\delta_{\rm ent}}
=\frac{\tilde{k}^2}{{k}^2}\frac{\tilde \delta_{\rm ent}}{\delta_{\rm ent}}\,.
 \end{eqnarray}
Thus the criterion is
\begin{eqnarray}
{\rm Case~I} &:~{t}_\mathrm{vir} < \tilde t_\mathrm{coll}\quad\Leftrightarrow\quad {k}^2\delta_{\rm ent}>\tilde{k}^2\tilde \delta_{\rm ent}\,,
\nonumber\\
{\rm Case~II}&:~{t}_\mathrm{vir} >\tilde t_\mathrm{coll}\quad\Leftrightarrow\quad {k}^2\delta_{\rm ent}<\tilde{k}^2\tilde \delta_{\rm ent}\,.
\label{scenario-criterion}
\end{eqnarray}

In both cases, once the velocity dispersion is shared over the whole patch of the PBH scale at $t=t_*$, 
the criterion for the PBH formation can be roughly evaluated by comparing the gravitational free-fall time 
$\Delta \tilde t_{\rm ff}\propto \tilde \rho^{-1/2}\propto \tilde R^{3/2}$ with the sound-wave crossing time $\Delta \tilde t_{\rm cross}\sim \tilde R/\sigma_v$, 
where $\tilde \rho$, $\tilde R$, and $\sigma_v$ are the density, the physical size of the scale $\tilde{k}$, and the velocity dispersion, respectively.
As the velocity dispersion increases,  the crossing time becomes shorter, and the PBH formation is eventually impeded.
The evaluation of the PBH formation condition is given in Sec.~\ref{sec:collapsecond}.

In Sec.~\ref{sec:threshold}, 
assuming that the scale $k$ can be treated independently of the scale $\tilde{k}$,
we consider a log-normal density spectrum $\ln \sigma(k)= \ln \sigma_0-\mu \ln (k/\tilde{k})^{2}$, where $\sigma_0$ and $\mu$ are dimensionless parameters, 
and evaluate the threshold amplitude of 
the density perturbation at horizon entry $\tilde \delta_{\rm ent}$.
In Sec.~\ref{sec:PBH_abundance}, the PBH production rate is roughly 
evaluated following the procedure provided in Ref.~\cite{Harada:2017fjm}, 
and the effect of the velocity dispersion is 
compared with the spin effects reported in Ref.~\cite{Harada:2017fjm}.

\section{Effect of the Velocity Dispersion against Collapse}
\label{sec:collapsecond}

The velocity dispersion may impede the gravitational collapse after the velocity dispersion is shared in the whole region of the PBH scale $\tilde{k}$. 
First, let us evaluate the impeding effect of the uniformly distributed velocity dispersion. We consider a spherical ball on the PBH scale.
Once the velocity dispersion is shared over the collapsing ball, 
the velocity dispersion simply increases as $\sigma_v \propto \tilde \rho^{1/3}$ as the system collapses adiabatically, according to Liouville's theorem.  
More explicitly, we have
\begin{align}
\sigma_v (t) =& \sigma_v (t_*) \left( \frac{\tilde \rho(t)}{\tilde \rho(t_*)} \right)^{1/3}=\sigma_v (t_*)  \frac{\tilde R(t_*)}{\tilde R(t)} , \label{sigma_v(t)}
\end{align}
where $t_*<t$ and $t_*$ is the time when the velocity dispersion is shared over the whole collapsing region. 

The sound-wave crossing time across the region of the size $R(t)$ is estimated as
\begin{align}
\Delta \tilde t_\text{cross} \simeq & \frac{\tilde R(t)}{\sigma_v(t)}=\frac{3}{2\sigma_v(t_*)} \frac{\tilde R(t)^2}{\tilde R(t_*)\tilde R_{\rm ent}}\tilde t_{\rm ent},  
\end{align}
where $\tilde R_{\rm ent}$ is the radius at the horizon entry time $\tilde t_{\rm ent}$ 
satisfying $\tilde R_{\rm ent}=1/\tilde H_{\rm ent}=3\tilde t_{\rm ent}/2$ in the Einstein-de Sitter background
with $\tilde H_{\rm ent}$ being the Hubble parameter at the horizon entry. 
On the other hand, the free-fall time is estimated as 
\begin{align}
  \Delta \tilde t_\text{ff} \simeq \left( \frac{\tilde R(t)}{\tilde R_\text{ent}} \right)^{3/2} \tilde t_\text{ent}.
\end{align}
Noting that $\Delta \tilde t_{\rm cross}\propto \tilde R^2$ and $\Delta \tilde t_{\rm ff}\propto \tilde R^{3/2}$, and $\Delta \tilde t_{\rm cross}>\Delta \tilde t_{\rm ff}$ at the time of horizon entry,
the collapse halts when $\Delta \tilde t_{\rm ff}=\Delta \tilde t_{\rm cross}$ if the radius is larger than the Schwarzschild radius at that time.
Let us denote the value of $\tilde R(t)$ at $\Delta \tilde t_{\rm ff}(t) =\Delta \tilde t_{\rm cross} (t)$ by $\tilde R_{\rm halt}$.  We find
\begin{equation}
  \tilde R_{\rm halt}=\frac{4}{9}\sigma_v^2(t_*)\frac{\tilde R^2(t_*)}{\tilde R_{\rm ent}^2}\tilde R_{\rm ent}\,. 
\end{equation}
If this is larger than $2\tilde M$, i.e., $\tilde R_{\rm halt}>\tilde R_{\rm ent}=1/\tilde H_{\rm ent}=2\tilde M$, the collapse halts before the PBH formation. 
Conversely, a PBH forms if $\tilde R_{\rm halt}<2\tilde M$.
Thus, the condition for the PBH \emph{non}-formation can be rewritten in the following simple form.
\begin{equation}
  \sigma_v(t_*)\tilde R(t_*)>\frac{3}{2}\tilde R_{\rm ent}. 
  \label{eq:nonform}
\end{equation}


To understand how the velocity dispersion is shared over the whole region of the size $1/\tilde{k}$, 
we introduce a closed FLRW collapsing model and consider its perturbations as given in Appendix~\ref{sec:generation}. 
Deferring detailed explanations of the complicated manipulations, here we just refer to the results. 
According to the discussion in Appendix~\ref{sec:generation}, depending on whether the halo formation on the scale $k$ can be completed 
before the collapse of the scale $\tilde{k}$, we can rewrite Eq.~\eqref{eq:nonform}, i.e., PBH non-formation condition, as 
\begin{align}
  &\delta_{\rm ent}<\frac{\tilde{k}^4}{k^4}&\text{for  }~
   \text{Case I~: } \quad k^2 \delta_{\rm ent}>\tilde{k}^2\tilde \delta_{\rm ent}\,, \label{tilde_delta_ent_inequality_case_I} \\
  &\delta_{\rm ent}>\frac{k}{\tilde{k}}\tilde \delta_{\rm ent}^{5/2}&\text{for  }~ 
   \text{Case II: }\quad k^2 \delta_{\rm ent}<\tilde{k}^2 \tilde \delta_{\rm ent}\,, \label{tilde_delta_ent_inequality_case_II}
\end{align}
where we have ignored factors of order unity. The qualitative features of the above conditions can be understood as follows.  

For Case I, the virialization of the scale $k$ occurs before the collapse of the would-be PBH scale. 
The condition~(\ref{tilde_delta_ent_inequality_case_I}) is a bit counter-intuitive.  So let us see why it is the case.
Since the gravitational potential $\Psi$ is constant in time in the matter-dominated universe, 
and $\delta_{\rm ent}\sim \Psi$ at horizon crossing, the virial velocity dispersion is estimated as $\sigma_{v}^2\sim\Psi\sim\delta_{\rm ent}$ (see Eq.~\eqref{eq:sigvdel}).
 The virial density is 
 $\rho_{\rm vir} = \rho_{\rm ent} (a_{\rm ent} / a_{\rm vir})^3 = \rho_{\rm ent} (\delta_{\rm ent} / \delta_{\rm vir})^3 \sim \rho_{\rm ent} \delta_{\rm ent}^3$ since $\delta \propto a$ and $\delta_{\rm vir} = O(1)$. 
As the collapse of a ball of the scale $\tilde{k}$ proceeds, the density $\tilde \rho$ reaches $\rho_{\rm vir}$ at $t=t_*$,
and the velocity dispersion is released to the entire ball, i.e., $\tilde \rho(t_*)=\rho_{\rm vir}$.
Hence we obtain the relation,
\begin{eqnarray}
\delta_{\rm ent}&\sim&\left(\dfrac{\rho_{\rm vir}}{\rho_{\rm ent}}\right)^{1/3}
= \left(\dfrac{\rho_{\rm vir}}{\tilde \rho_{\rm ent}}\right)^{1/3} \left(\dfrac{\tilde \rho_{\rm ent}}{\rho_{\rm ent}}\right)^{1/3}
= \left(\dfrac{\tilde \rho(t_*)}{\tilde \rho_{\rm ent}}\right)^{1/3}\left(\dfrac{a_{\rm ent}}{\tilde a_{\rm ent}}\right)
\nonumber\\
&=&\dfrac{\tilde R_{\rm ent}}{\tilde R(t_*)}\dfrac{\tilde{k}^2}{k^2}\,.
\end{eqnarray}
Recalling that $\delta_{\rm ent}\sim\sigma_v^2$ (Eq.~\eqref{eq:sigvdel}), this implies
\begin{eqnarray}
	\sigma_v^2\dfrac{\tilde R(t_*)^2}{\tilde R_{\rm ent}^2}\sim \delta_{\rm ent}\times\dfrac{\tilde{k}^4}{k^4}\frac{1}{\delta_{\rm ent}^2}
	=\dfrac{\tilde{k}^4}{k^4}\frac{1}{\delta_{\rm ent}}\,.
\end{eqnarray}
With Eq.~(\ref{eq:nonform}), this gives the condition quoted in Eq.~(\ref{tilde_delta_ent_inequality_case_I}). 
Here we have ignored all the coefficients of $\mathcal O(1)$. As one can see from the above estimate, the reason why the condition gives an upper bound on
$\delta_{\rm ent}$ is because it would make the small-scale virialized halos so small and dense that the large-scale region would collapse to
a black hole before the velocity dispersion is shared over the entire region of the PBH scale $\tilde{k}$. More details are given in Appendix~\ref{sec:case1}.

Let us now turn to Case II where the velocity dispersion is shared by the whole PBH region before the perturbations on the small scale $k$ are virialized.
The analysis in this case is a bit more complicated as one has to consider the growth of the small-scale perturbation in the collapsing region of the scale $\tilde{k}$. 
The detailed derivation of the condition~(\ref{tilde_delta_ent_inequality_case_II}) is defered to Appendix~\ref{sec:case2}.
Here we only mention that the reason why it depends on $\tilde \delta_\mathrm{ent}$ is precisely because of 
this fact that the small-scale perturbation grows within the collapsing ball of the large scale.
The reason why the condition gives a lower bound on $\delta_\mathrm{ent}$ is simply because 
a larger  $\delta_\mathrm{ent}$ gives a larger velocity dispersion, and hence it is easier to impede the PBH formation.

\section{Threshold of the Density Perturbation}
\label{sec:threshold}

As the effect of the perturbation on the small scale $k$ is statistical, hereafter, 
we replace $\delta_{\rm ent}$ by its root-mean-square value $\sigma_{\delta,{\rm ent}}(k)$.
Then, the condition for impeding the PBH formation due to velocity dispersion
can be summarized as follows: 
\begin{align}
 & \sigma_{\delta,\rm ent}(k)<\left(\frac{k}{\tilde{k}}\right)^{-4} & \text{for }~ \text{Case I~:} \quad \sigma_{\delta,{\rm ent}}(k)>\tilde \delta_{\rm ent}\left(\frac{k}{\tilde{k}}\right)^{-2}, 
  \label{eq:case1}\\
 & \sigma_{\delta,{\rm ent}}(k)>\tilde \delta_{\rm ent}^{5/2}\left(\frac{k}{\tilde{k}}\right) & \text{for }~  \text{Case II:} \quad \sigma_{\delta,{\rm ent}}(k)<\tilde \delta_{\rm ent}\left(\frac{k}{\tilde{k}}\right)^{-2},
  \label{eq:case2}
\end{align}
where we remind that the tilded quantities are those associated with the PBH scale.
So only the range $k > \tilde{k}$ is meaningful.  

Before proceeding to the threshold analysis, we mention a possible caveat about when one considers a continuous perturbation spectrum.
Although we consider the case $k\to\tilde{k}$ in our analysis, since we treat the two scales $k$ and $\tilde{k}$ as separate and independent scales, 
the validity of our argument becomes questionable in the limit $k\to\tilde{k}$.
To be more precise, since we consider the effect of the small-scale perturbation on the dynamics of the PBH scale only through the velocity dispersion,
which is a statistical quantity by definition, $k$ must be sufficiently larger than $\tilde k$ to allow statistical interpretations of 
the small-scale velocity effect.
Therefore, to avoid complications, we introduce the minimum applicable value for $k$, 
$k\geq k_{\rm min}$ where $k_{\rm min}>\tilde{k}$. Although we never assign an explicit value to $k_{\rm min}$, 
we expect $k_{\rm min}/\tilde{k}=\mathcal O(10)$. 
We also mention that the introduction of $k_{\rm min}$ renders the condition derived below a \emph{sufficient condition for PBH non-formation} or 
a \emph{necessary condition for PBH formation}.
This should be kept in mind below.

\begin{figure}[thb!] 
  \begin{center}
  \includegraphics[width= 0.6 \columnwidth]{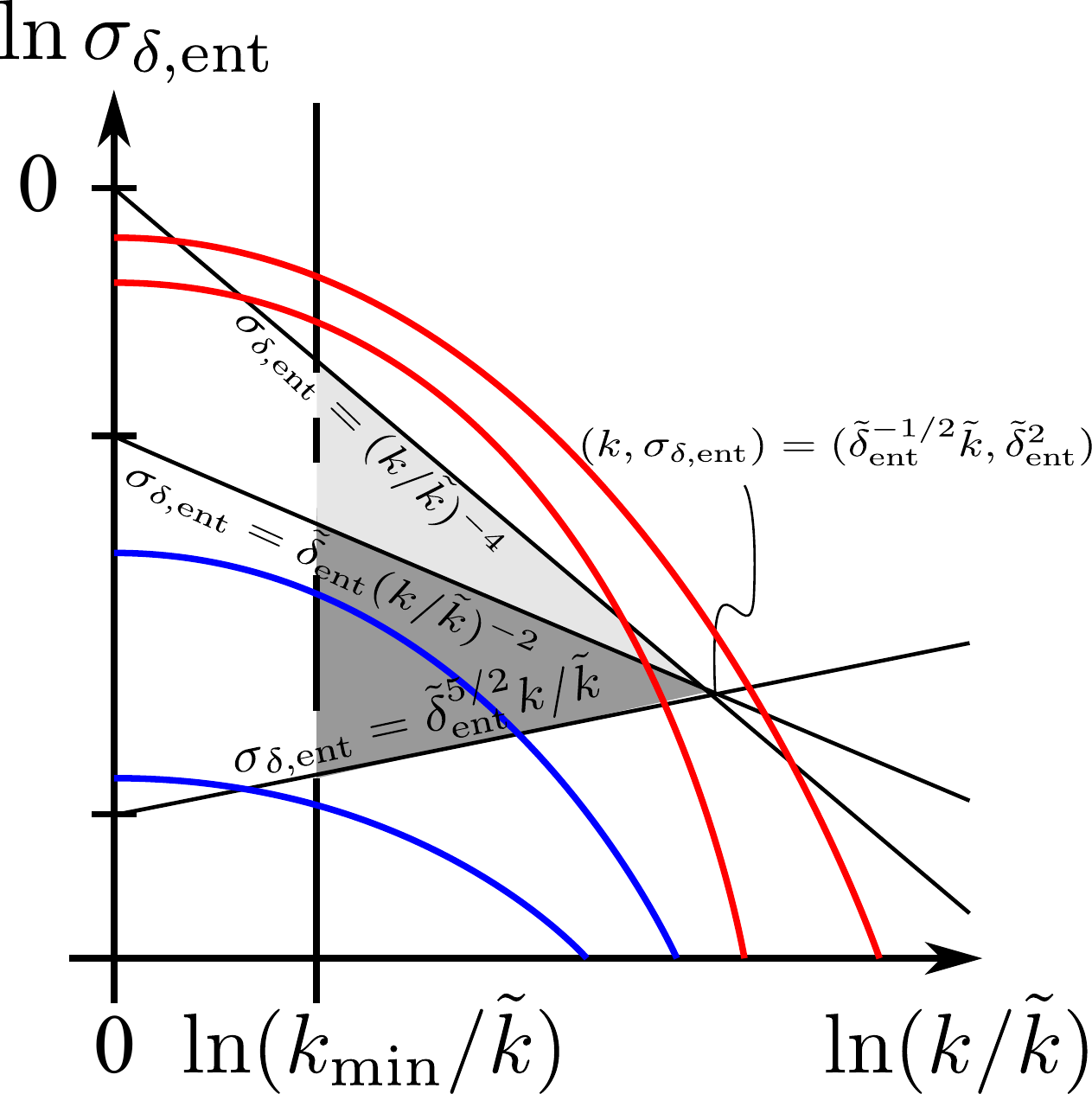}
\caption{The parameter space of PBH \emph{non}-formation in the $(k,\sigma_{\delta,\rm ent})$ plane.
  The shaded regions (Case I: light shaded, Case II: dark shaded) show the parameter region of PBH non-formation. 
The colored lines represent examples of density perturbation spectra peaked at the PBH scale.
 From top to bottom, the velocity dispersion does not impede the PBH formation (top red),
does do so (bottom red and top blue), and does not do so (bottom blue).
  }
  \label{fig:region}
  \end{center}
  \end{figure}

With this minimum value $k_{\rm min}$ for $k$ in mind, we schematically depict 
the region of PBH non-formation in Fig.~\ref{fig:region}, 
where light- and dark-shaded regions correspond to Case I and Case II, respectively.  
Once the density power spectrum $\sigma^2_{\delta, {\rm ent}}(k)$ is given as a function of $k$, and 
if the line describing the density power spectrum passes across the shaded region in Fig.~\ref{fig:region}, 
there exist small-scale perturbations that prevent a density perturbation of an
amplitude $\tilde \delta_{\rm ent}$ on the scale $\tilde{k}$ to form a PBH.
An important point in this diagram is the place where the lines representing 
the two conditions in (\ref{eq:case1}) and (\ref{eq:case2}) merge. It occurs at
$(k,\sigma_{\delta,\rm ent})=(\tilde\delta_{\rm ent}^{-1/2}\tilde{k},\tilde\delta_{\rm ent}^2)$.

Let us apply the above result to a couple of simple examples. 
When we consider the PBH formation, it is often assumed that there is a narrow peak 
in the power spectrum and abundant PBHs are produced on scales around the peak.
  In particular, as a comoving wavenumber $k$ and its horizon crossing
time $t$ during inflation is related as $k\sim {\mathrm e}^{Ht}$, a log-normal
distribution is naturally realized in many models of inflation.
  Then, around the peak of the power spectrum,  
the power spectrum may be well approximated by a log-normal spectrum:
  \begin{equation}
  \sigma^2_{\delta,{\rm ent}}(k)=\sigma^2_0 
\exp\left\{-2\mu \left[\ln \left(\frac{k}{\tilde{k}}\right)\right]^2\right\}, 
\end{equation}
or equivalently, 
\begin{equation}
  \ln \sigma_{\delta,{\rm ent}}(k)=-\mu \left[\ln \left(\frac{k}{\tilde{k}}\right)\right]^2+\ln \sigma_0, 
  \label{eq:spectrum}
\end{equation}
where $\mu$ is the non-dimensional parameter characterizing the width of the log-normal distribution.

\begin{figure}[thb!] 
  \begin{center}
    \includegraphics[width= 0.6 \columnwidth]{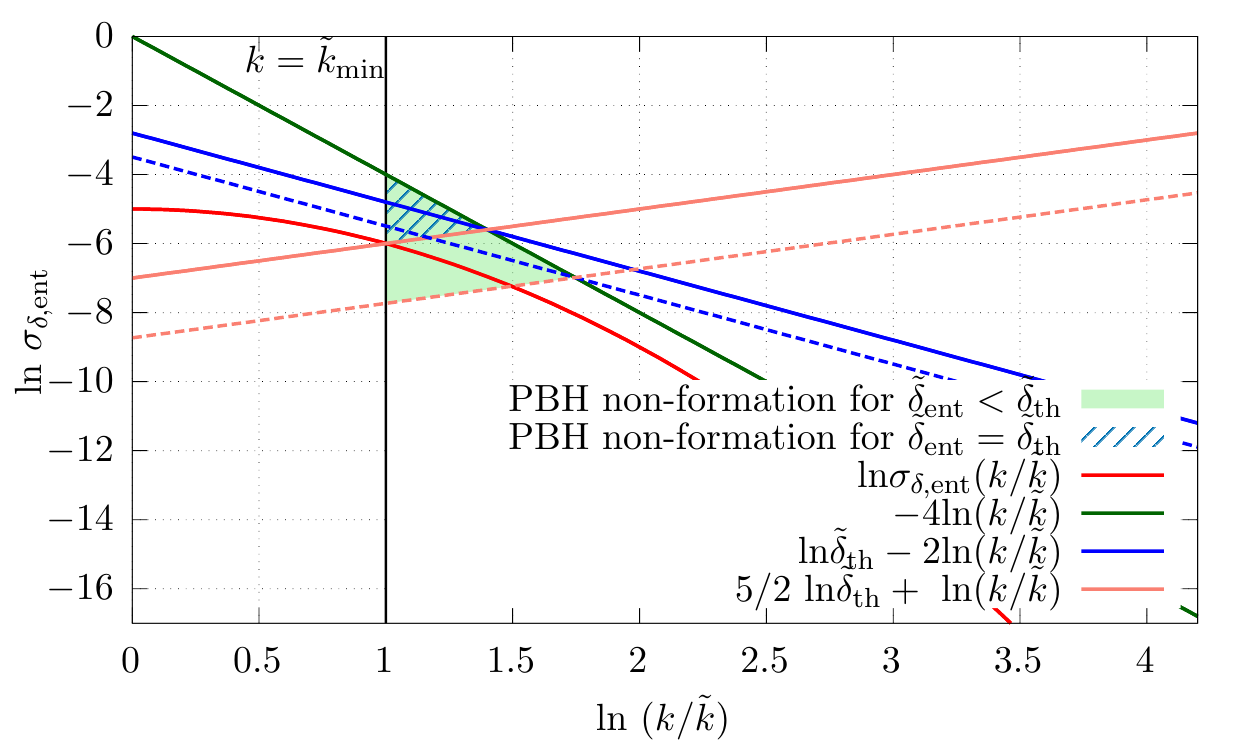}
    \includegraphics[width= 0.6 \columnwidth]{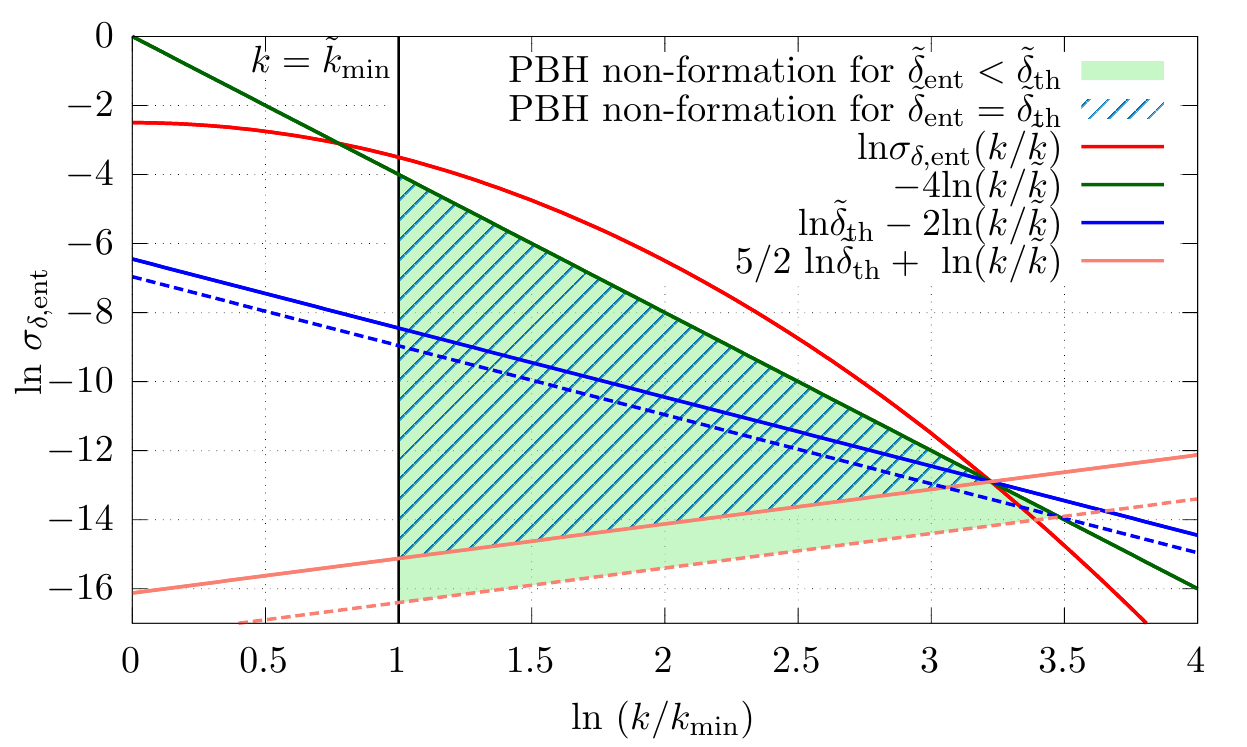}
    \caption{
      The region for PBH non-formation and the spectral line (red curve) for $\ln\sigma_0=-5$ (top panel) and $\ln\sigma_0=-2.5$ (bottom panel) with $\ln (k_{\rm min}/\tilde{k})=1$, $\mu=1$ are shown. The top (bottom) panel shows the case for $\sigma_0<\sigma_{\rm cr}(\mu,k_{\rm min})$ ($\sigma_0>\sigma_{\rm cr}(\mu,k_{\rm min})$), which corresponds to the blue (red) curves in Fig.~\ref{fig:region}. For $\tilde \delta_{\rm ent}<\tilde \delta_{\rm th}$ with $\tilde \delta_{\rm th}$ being defined in Eq. (\ref{eq:delta_th}), the lines for $\ln\sigma_{\delta,{\rm ent}}=\ln \tilde \delta_{\rm ent}-2\ln(k/\tilde k)$ and $\ln\sigma_{\delta,{\rm ent}}=5/2 \ln \tilde \delta_{\rm ent}+\ln(k/\tilde k)$ are shown by the blue and orange dashed lines, respectively.
      The green-shaded region is the region for PBH non-formation. Since the spectral line (red curve) passes through the green-shaded region, in this case, there exist small-scale perturbations that prevent the PBH formation on the scale $\tilde k$.  If $\tilde \delta_{\rm ent}$ is increased to the threshold value, i.e., $\tilde{\delta}_{\rm ent}=\tilde{\delta}_{\rm th}$, the blue and orange dashed lines get shifted up to the corresponding solid lines, so that the shaded triangle region of non-formation becomes narrower to the hatched triangle and touches the spectral line at a vertex.  
  }
  \label{fig:triangle1}
  \end{center}
  \end{figure}

First, let us consider the case when 
$\sigma_{\delta,{\rm ent}}(k_{\rm min})<(k_{\rm min}/\tilde{k})^{-4}$
 (blue lines in Fig.~\ref{fig:region}), namely, 
\begin{eqnarray}
  \ln \sigma_0<\ln \sigma_{\rm cr}(\mu,k_{\rm min})
\equiv\mu \left[\ln \left(\frac{k_{\rm min}}{\tilde{k}}\right)\right]^2
-4\ln \left(\frac{k_{\rm min}}{\tilde{k}}\right).  
\end{eqnarray}
In this case, 
since the spectrum must lie below the shaded region in Fig.~\ref{fig:region} to form PBHs,
we obtain the following inequality as the PBH formation condition 
(see the first panel in Fig.~\ref{fig:triangle1}):
\begin{eqnarray}
  &&\ln \left(\frac{k_{\rm min}}{\tilde{k}}\right)+\frac{5}{2}\ln \tilde \delta_{\rm ent}>\ln \sigma_{\delta,{\rm ent}}(k_{\rm min})\cr
  &\Leftrightarrow&
  \tilde \delta_{\rm ent}>
  \exp\left[-\frac{2}{5}\mu\left[\ln\left(\frac{k_{\rm min}}{\tilde{k}}\right)\right]^2\right]\left(\sigma_0 \frac{\tilde{k}}{k_{\rm min}}\right)^{2/5}. 
  \label{eq:th1}
\end{eqnarray}

In the case $\sigma_{\delta,{\rm ent}}(k_{\rm min})>(k_{\rm min}/\tilde{k})^{-4}$ 
(red lines in Fig.~\ref{fig:region}), namely, 
\begin{equation}
  \ln \sigma_0>\ln \sigma_{\rm cr}(\mu,k_{\rm min}), 
\end{equation}
the spectrum must lie above the shaded region to form PBHs. 
This condition is satisfied if the point
 $(k,\sigma_{\delta,\rm ent})=(\tilde\delta_{\rm ent}^{-1/2}\tilde{k},\tilde\delta_{\rm ent}^2)$
is below the spectral curve.
This leads to the condition,
\begin{eqnarray}
  2\ln \tilde \delta_{\rm ent}&<&
\ln \sigma_{\delta, {\rm ent}}\left(k=\tilde \delta_{\rm ent}^{-1/2}\tilde{k}\right)
=-\mu \left(\ln \tilde \delta_{\rm ent}^{-1/2}\right)^2+\ln \sigma_0\,. 
  \label{eq:forth2}
\end{eqnarray}
Noting that $\tilde\delta_{\rm ent}<1$, we find that the inequality \eqref{eq:forth2} 
implies the condition (see the second panel in Fig.~\ref{fig:triangle1}),
\begin{equation}
  \tilde \delta_{\rm ent}>
  \exp\left[-\frac{2}{\mu}\left(2+\sqrt{4+\mu\ln\sigma_0}\right)\right]. 
  \label{eq:th2}
\end{equation}

\begin{figure}[thb!] 
  \begin{center}
    \includegraphics[width= 0.8 \columnwidth]{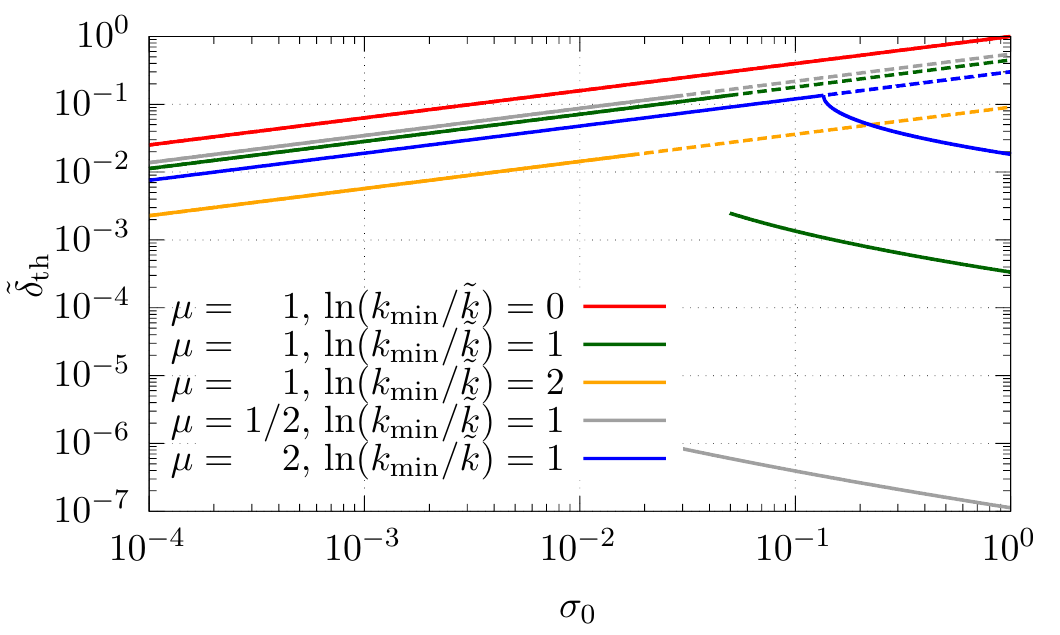}
    \caption{The density threshold $\tilde \delta_{\rm th}(\sigma_0,\mu,k_{\rm min})$ for PBH formation as a function of $\sigma_0$ for $(\mu,\ln(k_{\rm min}/\tilde{k}))=(1,0)$, $(1,1)$, $(1,2)$, $(1/2,1)$ and $(2,1)$. 
  }
  \label{fig:thresholds}
  \end{center}
  \end{figure}

To summarize, the \emph{necessary} condition for the PBH formation, \eqref{eq:th1} and \eqref{eq:th2}, is
\begin{equation}
\tilde \delta_{\rm ent}>
\tilde \delta_{\rm th}(\sigma_0,\mu,k_{\rm min})\equiv
  \left\{
  \begin{array}{ccl}
  \exp\left[-\dfrac{2}{5}\mu\ln^2\left({k_{\rm min}}/{\tilde{k}}\right)\right]
\left(\sigma_0({\tilde{k}}/{k_{\rm min}})\right)^{2/5}
  &; & \sigma_0<\sigma_{\rm cr}(\mu,k_{\rm min}), \\
  \exp\left[-\dfrac{2}{\mu}\left(2+\sqrt{4+\mu\ln\sigma_0}\right)\right]
  &; & \sigma_0>\sigma_{\rm cr}(\mu,k_{\rm min}).  
  \end{array}
  \right.
  \label{eq:delta_th}
 \end{equation}
We note that, as is explicitly shown in Fig.~\ref{fig:thresholds}, 
$\tilde \delta_{\rm th}$ is discontinuous at $\sigma_0= \sigma_{\rm cr}(\mu,k_{\rm min})$
as a function of $\sigma_0$ if 
\begin{equation}
  \left.\frac{d \ln \sigma_{\delta,{\rm ent}}(k)}{d\ln k}\right|_{k=k_{\rm min}}
>-4 \quad \Leftrightarrow  \quad \mu \ln \left(\frac{k_{\rm min}}{\tilde{k}}\right)<2\,.
\end{equation}

Some comments on the presence of the discontinuity in $\tilde\delta_{\rm th}$ are in order.  
If the root-mean-square value of perturbations on the small scale $\sigma_{\delta, \rm ent}(k)$ is large enough, 
the resulting virialized halos in a PBH scale region become so compact that they behave like particles
 at rest and never impede the PBH formation. 
On the other hand, if $\sigma_{\delta, \rm ent}(k)$ is very small, the velocity dispersion becomes
too small to affect the PBH formation. Thus the presence of the discontinuity reflects the fact 
that the velocity dispersion becomes an effective impeding factor only for a finite range 
of $\sigma_{\delta, \rm ent}(k)$. 
This results in the two types of spectra that may lead to the PBH formation, as shown by the top red and the bottom blue curves in Fig.~\ref{fig:region}.  
The transition from one type to the other corresponds to the discontinuity.  
On general grounds, however, one may expect that such a discontinuity would be smoothened when we improve the analyses.
For example, we required $\Delta \tilde t_\text{cross} < \Delta \tilde t_\text{ff}$ for the non-formation of PBHs,
 but the fate of would-be PBH over-dense regions may be probabilistic when $\Delta \tilde t_\text{cross} \simeq \Delta \tilde t_\text{ff}$.  
In addition, once we take into account the other impeding factors which we ignored in this paper, namely,
asphericity~\cite{Harada:2016mhb}, angular momentum~\cite{Harada:2017fjm}, and 
inhomogeneity~\cite{Khlopov:1980mg, Polnarev:1986, Kokubu:2018fxy}, the discontinuity will not play any role in determining
the PBH formation threshold for most of the parameter space of our interest (compare Figs.~\ref{fig:thresholds} and \ref{fig:beta}).

 For cosmological applications, one may expect that the width of the peak $\mu$ of the log-normal 
spectrum \eqref{eq:spectrum} would be of order unity, whereas the peak height $\sigma_0$ would not be
very close to unity since otherwise PBHs would be overproduced.  
Thus we expect that $\sigma_0<\sigma_{\rm cr}$ and the threshold in the first line of Eq.~(\ref{eq:delta_th}), 
$\tilde{\delta}_{\rm th}\propto \sigma_{0}^{2/5}$, or the condition from Case II, is likely to be relevant in such cases. 
We mention that if the peak is very sharp so that the width of the spectrum is much less than $k_{\rm min}$, namely, $\mu\ln^2(k_{\rm min}/\tilde{k})\gg 1$, 
the velocity dispersion becomes completely ineffective. 
In such a case, the other impeding factors we mentioned above will play a decisive role.

\section{PBH Formation Probability} 
\label{sec:PBH_abundance}

\begin{figure}[thbp] 
  \begin{center}
  \includegraphics[width= 0.7 \columnwidth]{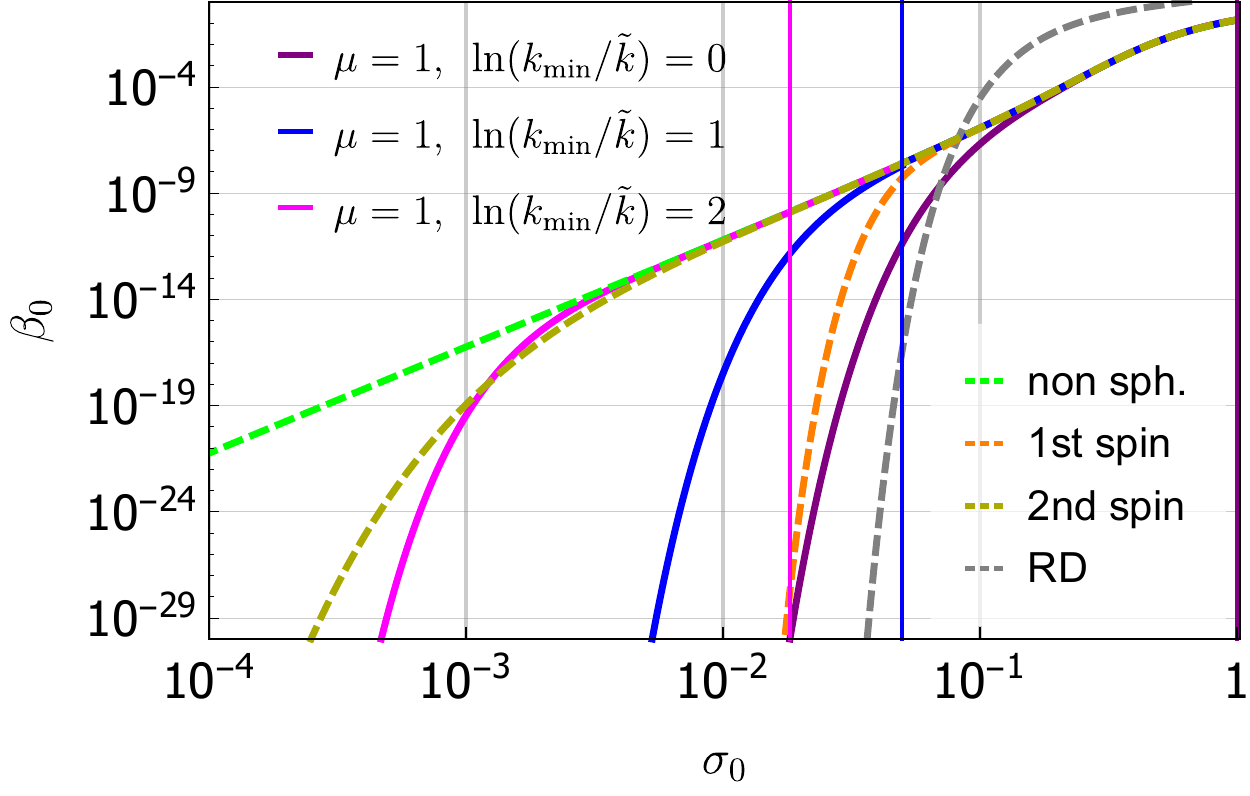}
  \vspace{5mm}

  \includegraphics[width= 0.7 \columnwidth]{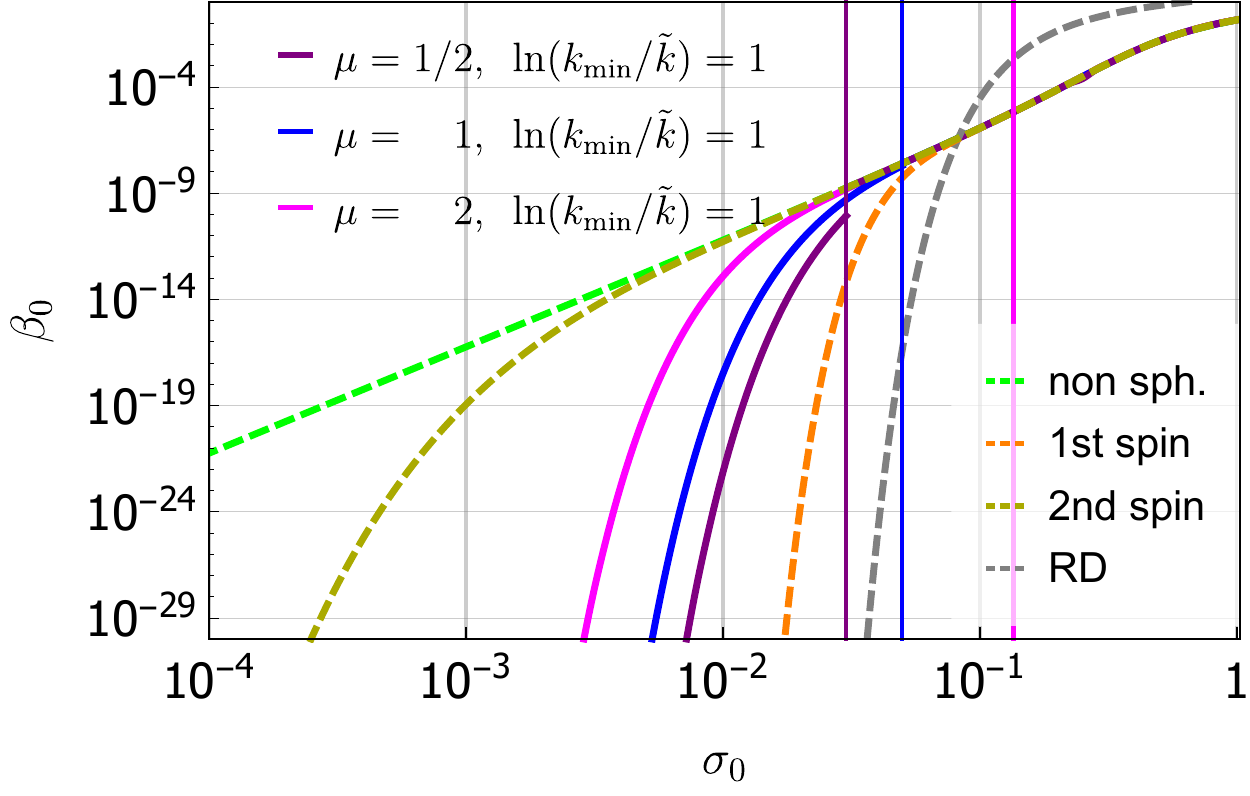}
  \caption{The dependence of the PBH production probability $\beta_0$ on the standard deviation $\sigma_0$ of the density perturbations at the PBH scale. 
The solid curves are our results including the effects of velocity dispersion 
for $(\mu,\ln(k_{\rm min}/\tilde{k}))=(1,0)$ (top panel; purple), $(1,1)$ (both panels; blue), $(1,2)$ (top panel; magenta), $(1/2,1)$ (bottom panel; purple),  and $(2,1)$ (bottom panel; magenta). 
The vertical solid lines indicate the corresponding values of $\sigma_{\rm cr}$.
For comparison, the first-order and second-order results in Ref.~\cite{Harada:2017fjm} are plotted as the orange (right) and dark orange (left) dashed lines, respectively. 
For these, parameters are chosen as $\mathcal{I} = 1$ and $q = \sqrt{2}$~\cite{Harada:2017fjm}. 
The common slope $\beta_0 \simeq 0.05556 \, \sigma_0^5$ represented by the green dashed line is the effect of anisotropy~\cite{Harada:2016mhb}. 
For comparison, the prediction of Carr's formula $\mathrm{Erfc} [ \tilde \delta_\text{th}/(\sqrt{2} \sigma_0) ]/2$ 
for the PBH production in the RD era is also shown by the dashed gray line with $\tilde \delta_\text{th} = 0.42$. 
}
  \label{fig:beta}
  \end{center}
  \end{figure}

In this section, we consider the PBH formation probability by taking into account not only the effect of velocity dispersion but also all the other 
impeding factors discussed previously in the literature.
First, let us recall the effect of angular momentum discussed in Ref.~\cite{Harada:2017fjm}. 
The threshold values due to the angular momentum estimated at first- and second-order perturbations 
are given in Eq.~(4.12) in Ref.~\cite{Harada:2017fjm} as 
\begin{eqnarray}
  \tilde \delta_{\rm th(1)}&=&\frac{3\cdot 2^2}{5^3}q^2, 
\label{eq:delta_th1}\\
  \tilde \delta_{\rm th(2)}&=&\left(\frac{2}{5}\mathcal I \sigma_0\right)^{2/3}, 
\label{eq:delta_th2}
\end{eqnarray}
where $q$ and  $\mathcal I$ are parameters of order unity describing the initial reduced quadrupole moment and the second-order
tidal contribution to the angular momentum, respectively,
and $\sigma_0$ is the standard deviation of the density perturbation at the PBH scale.
Assuming $q$ is independent of $\sigma_0$, the different dependence of the two thresholds on $\sigma_0$ leads to different shapes
of the PBH production probabilities as shown in Fig.~\ref{fig:beta}.

Second, let us turn to the effect of anisotropy. 
According to Ref.~\cite{Harada:2016mhb}, in an MD era, taking the anisotropy of the system into account, 
the PBH production probability $\beta_0$ is estimated as
\begin{align}
\beta_0 \simeq \int_0^\infty \mathrm{d}\alpha \int_{-\infty}^\alpha \mathrm{d} \beta \int_{- \infty}^\beta \mathrm{d} \gamma \, \Theta \left( \tilde \delta (\alpha, \beta, \gamma) - \delta_\text{th} \right) \Theta (1 - h (\alpha, \beta, \gamma)) w (\alpha, \beta, \gamma),  \label{eq:beta}
\end{align}
where $\alpha$, $\beta$, and $\gamma$ are the variables representing the anisotropy, $\Theta (\cdot )$ is the Heaviside step function,
 $\tilde \delta(\alpha, \beta, \gamma) = \alpha + \beta + \gamma$ is the density perturbation in terms of the anisotropy parameters, 
$h(\alpha, \beta, \gamma)$
is a function that determines the threshold based on the hoop conjecture~\cite{Klauder:1972lsv}, 
and $w (\alpha, \beta, \gamma)\propto
\exp \left[ -3\left( 2(\alpha^2 + \beta^2 + \gamma^2) - (\alpha \beta + \beta \gamma + \gamma \alpha) \right)/(2\sigma_0^2) \right]$ 
is the Doroshkevich probability distribution function~\cite{Doroshkevich:1970}. 
The result is that $\beta_0\propto \sigma_0^5$ with $\tilde \delta_{\rm th}=0$. See Refs.~\cite{Harada:2016mhb, Harada:2017fjm} for more details.

Finally, let us consider the effect of inhomogeneities discussed in \cite{Kokubu:2018fxy}. Their result indicates that $\beta_0 \propto \sigma^{3/2}$
if only the effect of inhomogeneities is taken into account. If we assume this impeding effect to work independently of the effect of anisotropy,
we would obtain $\beta_0\propto \sigma^5\times \sigma^{3/2}=\sigma^{13/2}$. 
However, as noted in  \cite{Kokubu:2018fxy}, the threshold
obtained there is a sufficient condition for the PBH formation. 
In other words, it is not a necessary condition for PBH formation. 
Therefore, to be conservative, we do not take its effect into account in this paper.

In Fig.~\ref{fig:beta}, we plot $\beta_0$ with $\tilde \delta_\text{th}$ determined by the consideration of the velocity dispersion using Eq.~\eqref{eq:delta_th}.  
For comparison, we also plot $\beta_0$ with $\tilde \delta_\text{th}$ determined by the consideration of the spin effects~\cite{Harada:2017fjm} using Eqs.~\eqref{eq:delta_th1} and \eqref{eq:delta_th2}. 
The common slope can be approximated as $\beta_0 \simeq 0.05556 \, \sigma_0^5$~\cite{Harada:2016mhb} (green dashed line).  
This is the effect of anisotropy. 
Different criteria on the density threshold result in different exponential falloffs. 
They can be understood by Carr's formula~\cite{Carr:1975qj}, according to which the PBH formation probability is roughly proportional 
to $\exp[-\tilde \delta_{\rm th}^2/(2 \sigma_0^2)]$. Applying this to our result \eqref{eq:delta_th} in the case  $\sigma_0<\sigma_{\rm cr}$,
$\tilde \delta_{\rm th}^2\propto \sigma_0^{4/5}$, the exponent is found to be proportional to $\sigma_0^{-6/5}$, as shown by the solid curves in Fig.~\ref{fig:beta}.  
On the other hand, the exponents are proportional to $\sigma_0^{-2}$ and $\sigma_0^{-2/3}$, respectively, for 
the first-order and second-order spin effects, respectively, 
as shown by the orange dashed (first-order) and dark orange dashed (second-order) curves.

From Fig.~\ref{fig:beta}, we see that the effect of the velocity dispersion can be the dominant factor against the PBH formation.  
Although the value of $\sigma_0$ where the probability exponentially drops significantly depends on $k_\mathrm{min}$, 
if the first-order spin effect is small, which is the case if $q\ll1$, or if $k_\mathrm{min} \simeq \tilde{k}$, the effect of the velocity dispersion is most significant. 
Here it should be noted that, in reality, all the impeding factors should be taken into account at the same time although we 
treated the effects of the spin and the velocity dispersion independently in this paper.  
\footnote{
From Fig.~\ref{fig:beta}, we find that the abundance produced in the RD epoch can be larger than the one in the MD at least for $\sigma_0 \gtrsim \mathcal{O}(0.1)$.
}

\section{Conclusions \label{sec:conclusion}}

In this paper, we discussed the effect of velocity dispersion on PBH formation.
For this purpose, we considered two distinct scales: One is a large scale that would form a PBH,
called the PBH scale, if there were no impeeding factors. 
The other is a small scale on which the velocity dispersion is generated. 
Then we derived the condition for the velocity dispersion on the small scale
to impede the PBH formation. 
Once the velocity dispersion is shared in the whole collapsing region, 
the diffusion associated with the velocity dispersion competes with 
the gravitational contraction. 
We obtained the simple criterion $\sigma_{v*}\tilde R_*<3\tilde M$ for the 
formation of a black hole of mass $\tilde M$ 
in terms of the size $\tilde R_*$ and the velocity dispersion $\sigma_{v*}$ 
at the time when the velocity dispersion is shared 
in the whole region~(see Eq.~\eqref{eq:nonform}). 
To apply this formula to the PBH formation, we considered two scenarios 
of the process. 

If the density perturbation on the small scale is sufficiently large, 
the virialization takes place before the collapsing time of the PBH scale. 
Then, the velocity dispersion in the virialized halos is released to the 
whole would-be PBH region when the mean density of that region becomes 
comparable to the virial density of halos, and it may impede PBH formation. 
However, if the density perturbation on the small scale is extremely large, 
the virialized halos become too compact and dense, and the dissolution of halos 
takes place too late to impede the PBH formation. 
Thus we obtain an upper bound for the amplitude of the 
small-scale density perturbation below which the PBH formation is impeded.

On the other hand, if the density perturbation on the small scale is extremely small,
the virialization can never be achieved before a PBH is formed.
However, even in this case, if the small-scale density perturbation is not too small,
it grows and may eventually enter the nonlinear regime. This growth of the small-scale
perturbation is enhanced by the fact that it behaves like a perturbation in
a closed collapsing background universe which the would-be PBH region mimics.
Then assuming that the velocity dispersion is generated and shared over the PBH scale 
when the amplitude of the small-scale density perturbation exceeds unity,
the PBH formation may be impeded. Thus, this scenario renders
the perturbation with a somewhat smaller amplitude to impede the PBH formation. 

The two conditions are summarized in the inequalities~\eqref{eq:case1} and \eqref{eq:case2}.  
These conditions can, in principle, apply to any functional form of the density power spectrum. 
Based on this result, we computed the PBH production rate, assuming that
the power spectrum is of a log-normal form, with the peak at the PBH scale.
The result is shown in Fig.~\ref{fig:beta}, where comparisons with the other impeding factors,
namely anisotropy and angular momentum, are also made.
We find that the effect of velocity dispersion can be a dominant factor 
to prevent PBH formation, although all these effects are subjected to various 
uncertainties, in addition to the spectral shape dependence,
 that affects the threshold value of the density perturbation amplitude.
It should be also noted that, if the spectrum has a sufficiently sharp peak around 
the PBH scale, the velocity dispersion is likely to be ineffective.  
  When we translate the observational constraints on the PBH abundance into
  those on model parameters in PBH formation scenarios, since the
  effect of the velocity dispersion may act as an additional factor to
  prevent PBH formation,  the allowed regions for the model parameters will 
  get wider~(see Figs.~10 and 11 in Ref.~\cite{Carr:2020gox} for the
  observational constraints on the PBH abundance).

One of the future directions is to properly include the effect of deviations from 
the idealized dust picture during the collapse. Besides the apparent fact that
the collapsing matter is not dust but collisionless cold particles in the simplest situation,
which naturally gives rise to the dispersion 
in the momentum space as well as in the real space,
if it interacts with each other or with other fields
 non-gravitationally, an additional pressure arises during virialization.  
If the matter is a massive scalar field, the quantum pressure due to the uncertainty principle
may play an essential role.
Recently, the PBH formation right after inflation when the inflaton is oscillating, 
which is effectively an MD era, has been discussed in 
Refs.~\cite{Cotner:2018vug,Kohri:2018qtx,Martin:2019nuw,Martin:2020fgl,Auclair:2020csm,Padilla:2021zgm,Eggemeier:2021smj,Hidalgo:2022yed}.  
It will be interesting to extend our general relativistic formalisms to these setups.

\section*{Acknowledgment}
The authors thank Yukawa Institute for Theoretical Physics at Kyoto University, where this work was initiated during the YITP-X-21-02  ``Brain-storming workshop on Primordial Black Holes and Gravitational Waves''.
This work was supported in part by IBS under the project code, IBS-R018-D1~(TT), and in part by 
JSPS KAKENHI Grant
Numbers JP19H01895~(CY, TH, MS), JP20H05850~(CY), JP20H05853~(CY, TH, MS), JP19K03876 (TH), JP17H01131 (KK), JP20H04750 (KK), JP22H05270 (KK), and JP20H04727 (MS).

\appendix

\section{Generation of Velocity Dispersion}
\label{sec:generation}

\subsection{Closed FLRW Collapsing Model}
\label{sec:closedFLRW}

To describe the evolution of over-dense regions, 
we introduce closed FLRW universes as models for the over-dense regions. 
We assume that the closed FLRW model can be applied to all over-dense regions on any scale, that is, they can apply to both scales of $\tilde{k}$ and $k$. 
Therefore, the equations given in this section can be applied to the scale $\tilde k$
replacing all the bare quantities~($a_{\rm ent}$, $\Omega_{\rm ent}$, $t_{\rm ent}$, $t_{\rm coll}$, $H_{\rm ent}$, $\delta_{\rm ent}$, $R_{\rm ent}$, $E$, $\mathcal K$ and $M$)  by the corresponding tilded quantities~($\tilde a_{\rm ent}$, $\tilde \Omega_{\rm ent}$, $\tilde t_{\rm ent}$, $\tilde t_{\rm coll}$, $\tilde H_{\rm ent}$, $\tilde \delta_{\rm ent}$, $\tilde R_{\rm ent}$, $\tilde E$, $\tilde{\mathcal K}$  and $\tilde M$). 

The Friedmann equation for a closed universe can be written in the following form:
\begin{align}
  \left( \frac{\dot{a}}{a} \right)^2 =& H_\text{ent}^2 \left [ \Omega_\text{ent} \left( \frac{a_\text{ent}}{a} \right)^3 + \left( 1 - \Omega_\text{ent} \right) \left( \frac{a_\text{ent}}{a} \right)^2 \right ],
  \end{align}  
  where $a$, $H_{\rm ent}$ and $\Omega_{\rm ent}$ are the scale factor, Hubble parameter, and cosmological density parameter at the horizon entry $t=t_{\rm ent}=2/(3 H_\text{ent})$, respectively. 
Note that, at this stage, the horizon entry time is just the time for reference because we have not introduced the scale of the over-dense region. 
The spatial curvature $\mathcal K$ is given by 
\begin{equation}
  \mathcal K:=(\Omega_{\rm ent}-1)H_{\rm ent}^2a_{\rm ent}^2. 
  \end{equation}
The parametric solution for the scale factor $a$ is 
\begin{align}
\frac{a}{a_\text{ent}} = & \frac{\Omega_\text{ent}}{2 (\Omega_\text{ent}-1)} (1 - \cos \theta), \\
t = & \frac{\Omega_\text{ent}}{2 H_\text{ent} (\Omega_\text{ent} -1 )^{3/2} } ( \theta - \sin \theta ),
\end{align}
where the parameter $\theta$ is related to the conformal time $\eta$ by $\theta=\sqrt{\mathcal K}\eta$.

Taking the Einstein-de Sitter~(EdS) universe as the background universe and considering the uniform Hubble slice, 
we can obtain the value of the density perturbation at the horizon entry as 
\begin{equation}
  \delta_{\rm ent}=\Omega_{\rm ent}-1, 
\end{equation}
where we have used the fact that the background density is given by the critical density $3H^2/(8\pi)$ for the uniform Hubble slice. 
Hereafter we assume $\delta_\text{ent} \ll 1$, which can be satisfied even for a rarely high-density region leading to PBH formation in an MD era. 

For $\theta=\sqrt{\mathcal K}\eta\ll1$,  we find $a\propto t^{2/3}$ at the leading order, and the over-dense region can be approximated by 
an EdS universe. 
More specifically, we find 
\begin{align}
  \frac{a}{a_\text{ent}} =& \frac{\Omega_{\rm ent}}{4(\Omega_{\rm ent}-1)}\mathcal K \eta^2+\frac{1}{\Omega_{\rm ent}-1}\mathcal O\left[(\mathcal K\eta^2)^2\right], \\ 
  t =& \frac{\Omega_{\rm ent}a_{\rm ent}\eta}{12(\Omega_{\rm ent}-1)}\mathcal K\eta^2+\frac{a_{\rm ent}\eta}{\Omega_{\rm ent}-1}\mathcal O\left[(\mathcal K\eta^2)^2\right].
  \label{eq:backt}
  \end{align}  
Since the value of $\eta$ at the horizon entry is given by $\eta_{\rm ent}\sim 1/(H_{\rm ent}a_{\rm ent})\sim \delta_{\rm ent}/\sqrt{\mathcal K}$, 
we find $\eta<\eta_{\rm ent}\ll1/\sqrt{\mathcal K}$ before the horizon entry, and then 
the EdS approximation is valid before the horizon entry. 
The time at the maximum expansion~($\theta=\pi$), $t_\text{max}$, and 
the collapsing time~($\theta=2\pi$), $t_\text{coll}$, are given by 
\begin{align}
t_\text{coll}=2t_{\rm max}= \frac{\pi\Omega_{\rm ent}}{H_\text{ent} (\Omega_{\rm ent}-1)^{3/2}}\simeq & \frac{\pi}{H_\text{ent} \delta_\text{ent}^{3/2}}.
\label{eq:tcoll}
\end{align}
At the maximum expansion, the radius of the over-dense region, $R_\text{max}$, becomes 
\begin{equation}
  R_\text{max} \simeq \frac{1}{H_\text{ent} \delta_\text{ent}} = \frac{R_\text{ent} }{ \delta_\text{ent}}. 
  \label{eq:RmaxRent}
\end{equation}
At this moment, the energy of the over-dense region can be parametrized as 
\begin{equation}
  E = - h \frac{M^2 }{ R_\text{max}}
\end{equation}
with $M= 1/(2H_\text{ent})$ being the horizon mass at the horizon entry and $h$ is an $\mathcal{O}(1)$ dimensionless parameter. 
For a uniform ball, $h = 3/5$.


\subsection{Virialization}

After the collapse, if a black hole does not form, the collapsed region is expected to be virialized in a time scale comparable to the collapsing time,
\begin{align}
t_\text{vir} = s t_\text{coll},
\end{align}
with $s = \mathcal{O}(1) > 1$ being another dimensionless parameter.
Let us consider the generation process of the velocity dispersion and how it can be shared in  
the whole region of the scale $\tilde{k}$. 
As is stated in Sec.~\ref{sec:overview}, here we consider two possible scenarios 
depending on whether the virialization of the scale $k$ can be realized before $t=\tilde t_{\rm coll}$ or not. 
The condition for the realization of the virialization is given by 
\begin{equation}
  t_{\rm vir}=s t_{\rm coll}<\tilde t_{\rm coll}. 
\end{equation}
By using Eq.~\eqref{eq:tcoll}, 
the inequality can be rewritten as 
\begin{align}
 &\qquad s H_{\rm ent}^{-1}\delta_{\rm ent}^{-3/2}<\tilde H_{\rm ent}^{-1} \tilde \delta_{\rm ent}^{-3/2} \nonumber \\
 \Leftrightarrow &\qquad s k^{-3}\delta_{\rm ent}^{-3/2}< \tilde{k}^{-3}\tilde \delta_{\rm ent}^{-3/2} \nonumber \\
 \Leftrightarrow &\qquad \frac{k^2}{\tilde{k}^2}\frac{\delta_{\rm ent}}{\tilde \delta_{\rm ent}}>s^{2/3}=\mathcal O(1), 
  \label{eq:condvir}
\end{align}
where we have used the relation $\tilde H_{\rm ent}/H_{\rm ent}=\tilde{k}^3/k^3$. 
Hereafter we individually discuss the two cases $t_{\rm vir}<\tilde t_{\rm coll}$~(case I) and $t_{\rm vir}>\tilde t_{\rm coll}$~(case II) in subsections \ref{sec:case1} and \ref{sec:case2}, respectively.

\subsection{Case I: Generation through the Virialization}
\label{sec:case1}

Here, we consider the case $t_{\rm vir}=st_{\rm coll}<\tilde t_{\rm coll}$, namely, the condition \eqref{eq:condvir} is satisfied.  
The virial theorem states that the kinetic energy $K$ and the potential energy $U$ satisfies $K=-U/2=h {M^2}/{R_{\rm vir}}$, 
where $R_{\rm vir}$ is the radius of the virialized halo. 
Since the total energy is conserved and is given by $-h{M^2}/{R_{\rm max}}$ at the maximum expansion, 
we obtain 
\begin{equation}
  E=K+U=-K= -\frac{1}{2}M \sigma_v^2=-h\frac{M^2}{R_{\rm max}}. 
\end{equation}
From this equation, we find
\begin{equation}
  \sigma_v^2=2h\frac{M}{R_{\rm max}}=h\frac{2M}{R_{\rm ent}}\frac{R_{\rm ent}}{R_{\rm max}}\simeq h \delta_{\rm ent}, 
  \label{eq:sigvdel}
\end{equation}
where we used $2M=R_{\rm ent}$ and Eq.~\eqref{eq:RmaxRent}. 
In addition, from $E=U/2=U_{\rm max}$, we find $R_{\rm vir}=R_{\rm max}/2$, or equivalently, 
\begin{equation}
  \rho_{\rm vir}=8\rho_{\rm max}\simeq8\rho_{\rm ent}\delta_{\rm ent}^3. 
\end{equation}

The virialized halos behave as macroscopic particles until the mean density of the scale $\tilde{k}$ becomes comparable to the virial density $\rho_{\rm vir}$. 
Thus, we identify the time $t=t_*$ by $\tilde \rho(t_*)=\rho_{\rm vir}=8\rho_{\rm ent}\delta_{\rm ent}^3$. 
This equation can be rewritten as 
\begin{equation}
  2\delta_{\rm ent}=\left(\frac{\tilde \rho(t_*)}{\rho_{\rm ent}}\right)^{1/3}=\frac{\tilde R(t_{\rm ent})}{\tilde R(t_*)}=\frac{\tilde R_{\rm ent}}{\tilde R(t_*)}\frac{a_{\rm ent}}{\tilde a_{\rm ent}}
  =\frac{\tilde R_{\rm ent}}{\tilde R(t_*)}\frac{\tilde{k}^2}{k^2}. 
\end{equation}
From this equation and $\sigma_v^2(t_*)\simeq h \delta_{\rm ent}$, we obtain 
\begin{equation}
  \sigma_v^2(t_*)\tilde R^2(t_*)\simeq\frac{h}{4\delta_{\rm ent}}\frac{\tilde{k}^4}{k^4}\tilde R_{\rm ent}^2. 
\end{equation}
Then, the condition \eqref{eq:nonform} can be reduced to 
\begin{equation}
  \delta_{\rm ent}<\frac{h}{9}\frac{\tilde{k}^4}{k^4}. 
\end{equation}
At first glance, 
the last inequality seems counterintuitive because we need a smaller value of $\delta_{\rm ent}$ to 
obtain the significant effect of the velocity dispersion originating from the smaller scale perturbations. 
However, this inequality indeed makes sense as follows. 
Since the virial radius is given by $R_{\rm vir}=R_{\rm ent}
/({2\delta_{\rm ent}})$, 
the larger $\delta_{\rm ent}$ is, the smaller the size of the virialized halos becomes. 
Therefore, under the condition that the virialization is realized before $t=\tilde t_{\rm coll}$, 
if the value of $\delta_{\rm ent}$ is larger than the critical value, 
the virialized halos are too small and the halo dissolution is too late to 
prevent PBH formation.

\subsection{Case II: Direct Generation from nonlinear Perturbations}
\label{sec:case2}

Here we consider the case $t_{\rm vir}=st_{\rm coll}>\tilde t_{\rm coll}$, namely, the following inequality is satisfied: 
\begin{equation}
  \frac{k^2}{\tilde{k}^2}\frac{\delta_{\rm ent}}{\tilde \delta_{\rm ent}}>s^{2/3}=\mathcal O(1). 
  \label{eq:condnonvir}
\end{equation} 
Even in this case, the growth of the smaller-scale perturbations is promoted as the contraction of the scale $\tilde{k}$ proceeds. 
Then, the smaller-scale perturbations may get into the nonlinear regime and produce caustics, and velocity dispersion is generated before the collapse of the scale $\tilde{k}$. 
We assume that the velocity dispersion is shared in the whole region of the scale $\tilde{k}$ at the same time $t_*$ of its generation. 
To quantitatively evaluate the value of the velocity dispersion, let us consider the 
linear-perturbation equation in the background closed universe. 

\subsubsection{Density perturbations in the closed FLRW universe}

In the comoving gauge, the equation for the density perturbations of the scale $\tilde{k}$ in the closed universe is given by
(see, e.g., Ref.~\cite{Weinberg:1972kfs})
\begin{equation}
(1-\cos\theta)\frac{\mathrm d^2}{\mathrm d\theta^2}\delta_{\rm co}+\sin\theta\frac{\mathrm d}{\mathrm d\theta}\delta_{\rm co}-3\delta_{\rm co}=0. 
\end{equation}
The growing-mode solution is given as follows (see, e.g., Ref.~\cite{Weinberg:1972kfs}): 
\begin{equation}
\delta_{\rm co}=\frac{\delta_{\rm co,max}}{2}\left(\frac{5+\cos\theta}{1-\cos\theta}-\frac{3\theta\sin\theta}{(1-\cos\theta)^2}\right)
=\delta_{\rm co,max}\left(\frac{\theta^2}{20}+\mathcal O(\theta^4)\right), 
\label{eq:delcosmth}
\end{equation}
where $\delta_{\rm co,max}=\delta_{\rm co}(\pi)$. 

In order to relate the evolution of the perturbation $\delta_{\rm co}$ to the density perturbation $\delta$ 
defined in the flat FLRW model with the uniform Hubble slice, 
we need to clarify the relation between the values of $\delta_{\rm ent}$ and $\delta_{\rm co,max}$. 
For this purpose, let us consider the value of $\delta_{\rm co}$ at the horizon entry. 
The horizon entry of the scale $k$ is earlier than that of the scale $\tilde{k}$,  so the EdS approximation is valid at the horizon entry time 
as is discussed in Sec.~\ref{sec:closedFLRW}. 
Then, we obtain 
\begin{eqnarray}
&&a_{\rm ent}H_{\rm ent}=k, 
\\
&&H_{\rm ent}=\tilde H_{\rm ent}\frac{k^3}{\tilde{k}^3}=\frac{2}{3t_{\rm ent}}. 
\end{eqnarray}
From the second equation, we find 
\begin{eqnarray}
t_{\rm ent}&=&\frac{2}{3\tilde H_{\rm ent}}\frac{\tilde{k}^3}{k^3}. 
\end{eqnarray}

Let us 
express $t_{\rm ent}$ by using the conformal time from Eq.~\eqref{eq:backt} as follows: 
\begin{equation}
t_{\rm ent}\simeq\frac{1}{12\tilde H_{\rm ent}}\left(\tilde H_{\rm ent}\tilde a_{\rm ent}\eta_{\rm ent}\right)^3, 
\end{equation}
where we have used the approximation $\tilde \Omega_{\rm ent}\simeq 1$. 
Then we obtain 
\begin{equation}
\eta_{\rm ent}=2\frac{\tilde{k}}{k}\frac{1}{\tilde a_{\rm ent}\tilde H_{\rm ent}}. 
\end{equation}
From Eq.~\eqref{eq:delcosmth}, we can find 
\begin{equation}
\delta_{\rm co,ent}\simeq \tilde{\mathcal K} \eta_{\rm ent}^2\frac{\delta_{\rm co,max}}{20}=\frac{1}{5}(\tilde \Omega_{\rm ent}-1)\frac{\tilde{k}^2}{k^2}\delta_{\rm co,max}
=\frac{1}{5}\frac{\tilde{k}^2}{k^2}\tilde \delta_{\rm ent}\delta_{\rm co,max}. 
\end{equation}
Rewriting $\delta_{\rm co,max}$ by using $\delta_{\rm co,ent}$, we obtain 
\begin{equation}
\delta_{\rm co}=\frac{5}{2}\frac{k^2}{\tilde{k}^2}\frac{\delta_{\rm co,ent}}{\tilde \delta_{\rm ent}}
\left(\frac{5+\cos\theta}{1-\cos\theta}-\frac{3\theta\sin\theta}{(1-\cos\theta)^2}\right). 
\end{equation}
It is known that the density perturbations $\delta_{\rm co}$ in the comoving gauge and $\delta$ in the uniform Hubble gauge 
is related to each other as $\delta_{\rm co}\simeq ({3}/{5})\delta$ on super-horizon scales (see, e.g., Ref.~\cite{Harada:2015yda}). 
Then, we extrapolate this relation to the horizon entry, that is, we 
use the relation $\delta_{\rm co,ent}\simeq  ({3}/{5})\delta_{\rm ent}$. 
By using this relation, we finally obtain the following expression: 
\begin{equation}
\delta_{\rm co}=\frac{3}{2}\frac{k^2}{\tilde{k}^2}\frac{\delta_{\rm ent}}{\tilde \delta_{\rm ent}}
\left(\frac{5+\cos\theta}{1-\cos\theta}-\frac{3\theta\sin\theta}{(1-\cos\theta)^2}\right). 
\end{equation}

\subsubsection{Estimation of the velocity dispersion}

To evaluate the velocity dispersion, 
we consider the velocity perturbation in the comoving gauge. 
In the comoving gauge, the velocity potential $\phi_v$ is related to the density perturbation as \cite{Weinberg:1972kfs}
\begin{equation}
\phi_v=-(\triangle +3\tilde{\mathcal K})^{-1}\frac{\mathrm d}{\mathrm d\eta} \delta_{\rm co}, 
\end{equation}
where $\triangle$ is the Laplacian. 
The spatial curvature contribution can be estimated as 
$\tilde{\mathcal K}=\tilde \delta_{\rm ent}\tilde{k}^2\ll k^2$, and the contribution is negligible compared to the Laplacian term. 
Then, we have 
\begin{equation}
\phi_v\simeq k^{-2}\frac{\mathrm d}{\mathrm d\eta} {\delta}_{\rm co}. 
\end{equation}
The absolute value $v$ of the velocity can be estimated as follows: 
\begin{equation}
v \simeq {k} \phi_v\simeq k^{-1}\frac{d}{d\eta} 
{\delta}_{\rm co}. 
\label{eq:sigv1}
\end{equation}
From the following equations 
\begin{eqnarray}
\tilde R(\theta)&=&\tilde R_{\rm ent}\frac{\tilde a}{\tilde a_{\rm ent}}=\frac{\tilde R_{\rm ent}}{2}\frac{\tilde \Omega_{\rm ent}}{\tilde \Omega_{\rm ent}-1}(1-\cos\theta), \\
v(\theta)&\simeq&\frac{\frac{\mathrm d}{\mathrm d\eta}  {\delta}_{\rm co}}{k}=\frac{9}{2}\frac{k}{\tilde{k}}\frac{1}{\sqrt{\tilde \Omega_{\rm ent}-1}}\delta_{\rm ent}\frac{2\theta+\theta\cos\theta-3\sin\theta}{(1-\cos\theta)^2}, 
\end{eqnarray}
we obtain 
\begin{equation}
\frac{\tilde R(\theta)}{\tilde R_{\rm ent}}v(\theta)\simeq\frac{9}{4}\frac{k}{\tilde{k}}\tilde \delta_{\rm ent}^{-3/2}\delta_{\rm ent}
\frac{2\theta+\theta\cos\theta-3\sin\theta}{1-\cos\theta},  
\label{eq:rv}
\end{equation}
where we have used $\tilde \Omega_{\rm ent}-1=\tilde \delta_{\rm ent}$.

Let us suppose that the linear perturbations continue to grow until they enter the nonlinear regime, 
and caustics appear. 
Assuming this happens around when $\delta_{\rm co}\sim1$, we adopt the 
value of $v$ at the time $t=t_*$ defined by $\delta_{\rm co}=1$ for the value of $\sigma_v(t_*)$. 
Letting $\theta_*$ denote the value of $\theta$ at $t=t_*$, we find 
\begin{equation}
1=\delta_{\rm co}(\theta_*)=\frac{3}{2}\frac{k^2}{\tilde{k}^2}\frac{\delta_{\rm ent}}{\tilde \delta_{\rm ent}}
\frac{5-4\cos\theta_*-\cos^2\theta_*-3\theta_*\sin\theta_*}{(1-\cos\theta_*)^2}. 
\end{equation}
From the inequality \eqref{eq:condnonvir}, we can find $\theta_*\gtrsim 2.2$. 
Arranging the equation for $\theta_*$, we can derive the following equation: 
\begin{equation}
  (1-\cos\theta_*)^{-1}=2^{2/3}3^{-2/3}\tilde \delta_{\rm ent}^{2/3}\left(\frac{k}{\tilde{k}}\right)^{-4/3}\delta_{\rm ent}^{-2/3}
  \left(\frac{5-4\cos\theta_*-\cos^2\theta_*-3\theta_*\sin\theta_*}{\sqrt{1-\cos\theta_*}}\right)^{-2/3}. 
  \end{equation}
Substituting this equation into Eq.~\eqref{eq:rv}, we obtain
\begin{eqnarray}
&&\frac{\tilde R(\theta_*)}{\tilde R_{\rm ent}}\sigma_v(\theta_*)\simeq\frac{\tilde R(\theta_*)}{\tilde R_{\rm ent}}v(\theta_*)\simeq3^{4/3}2^{-4/3}\left(\frac{k}{\tilde{k}}\right)^{-1/3}\tilde \delta_{\rm ent}^{-5/6}
\delta_{\rm ent}^{1/3}\Xi(\theta_*), 
\end{eqnarray}
where 
\begin{eqnarray}
\Xi(\theta_*):=(2\theta_*+\theta_*\cos\theta_*-3\sin\theta_*)\left(\frac{5-4\cos\theta_*-\cos^2\theta_*-3\theta_*\sin\theta_*}{\sqrt{1-\cos\theta_*}}\right)^{-2/3}. 
\end{eqnarray}
Because $\Xi(\theta_*)$ is a monotonic function of $\theta_*$ satisfying $\Xi(2.2)\simeq 0.56$ and $\Xi(2\pi)=(3\pi)^{1/3}\simeq 2.1$, 
we can treat $\Xi(\theta_*)$ as a factor of order 1. 
Then the PBH non-formation condition \eqref{eq:nonform} can be rewritten as 
 \begin{equation}
 b\frac{\tilde{k}}{k}\tilde \delta_{\rm ent}^{-5/2}\delta_{\rm ent}>1, 
 \end{equation}
 where 
 \begin{equation}
 b:=\frac{3}{2}\Xi^3(\theta_*)=\mathcal O(1). 
 \end{equation}

 \providecommand{\href}[2]{#2}\begingroup\raggedright\endgroup


\begin{thebibliography}{100}

  \bibitem{Ghez:2008ms}
  A.~M. Ghez {\em et~al.}, ``{Measuring Distance and Properties of the Milky
    Way's Central Supermassive Black Hole with Stellar Orbits},''
    \href{http://dx.doi.org/10.1086/592738}{{\em Astrophys. J.} {\bfseries 689}
    (2008) 1044--1062}, \href{http://arxiv.org/abs/0808.2870}{{\ttfamily
    arXiv:0808.2870 [astro-ph]}}.
  
  \bibitem{Gillessen:2009ht}
  S.~Gillessen, F.~Eisenhauer, T.~K. Fritz, H.~Bartko, K.~Dodds-Eden, O.~Pfuhl,
    T.~Ott, and R.~Genzel, ``{The orbit of the star S2 around SgrA* from VLT and
    Keck data},'' \href{http://dx.doi.org/10.1088/0004-637X/707/2/L114}{{\em
    Astrophys. J. Lett.} {\bfseries 707} (2009) L114--L117},
    \href{http://arxiv.org/abs/0910.3069}{{\ttfamily arXiv:0910.3069
    [astro-ph.GA]}}.
  
  \bibitem{Genzel:2010zy}
  R.~Genzel, F.~Eisenhauer, and S.~Gillessen, ``{The Galactic Center Massive
    Black Hole and Nuclear Star Cluster},''
    \href{http://dx.doi.org/10.1103/RevModPhys.82.3121}{{\em Rev. Mod. Phys.}
    {\bfseries 82} (2010) 3121--3195},
    \href{http://arxiv.org/abs/1006.0064}{{\ttfamily arXiv:1006.0064
    [astro-ph.GA]}}.
  
  \bibitem{GRAVITY:2018ofz}
  {\bfseries GRAVITY} Collaboration, R.~Abuter {\em et~al.}, ``{Detection of the
    gravitational redshift in the orbit of the star S2 near the Galactic centre
    massive black hole},''
    \href{http://dx.doi.org/10.1051/0004-6361/201833718}{{\em Astron. Astrophys.}
    {\bfseries 615} (2018) L15},
    \href{http://arxiv.org/abs/1807.09409}{{\ttfamily arXiv:1807.09409
    [astro-ph.GA]}}.
  
  \bibitem{LIGOScientific:2016aoc}
  {\bfseries LIGO Scientific, Virgo} Collaboration, B.~P. Abbott {\em et~al.},
    ``{Observation of Gravitational Waves from a Binary Black Hole Merger},''
    \href{http://dx.doi.org/10.1103/PhysRevLett.116.061102}{{\em Phys. Rev.
    Lett.} {\bfseries 116} no.~6, (2016) 061102},
    \href{http://arxiv.org/abs/1602.03837}{{\ttfamily arXiv:1602.03837 [gr-qc]}}.
  
  \bibitem{LIGOScientific:2018mvr}
  {\bfseries LIGO Scientific, Virgo} Collaboration, B.~P. Abbott {\em et~al.},
    ``{GWTC-1: A Gravitational-Wave Transient Catalog of Compact Binary Mergers
    Observed by LIGO and Virgo during the First and Second Observing Runs},''
    \href{http://dx.doi.org/10.1103/PhysRevX.9.031040}{{\em Phys. Rev. X}
    {\bfseries 9} no.~3, (2019) 031040},
    \href{http://arxiv.org/abs/1811.12907}{{\ttfamily arXiv:1811.12907
    [astro-ph.HE]}}.
  
  \bibitem{LIGOScientific:2021usb}
  {\bfseries LIGO Scientific, VIRGO} Collaboration, R.~Abbott {\em et~al.},
    ``{GWTC-2.1: Deep Extended Catalog of Compact Binary Coalescences Observed by
    LIGO and Virgo During the First Half of the Third Observing Run},''
    \href{http://arxiv.org/abs/2108.01045}{{\ttfamily arXiv:2108.01045 [gr-qc]}}.
  
  \bibitem{EventHorizonTelescope:2019dse}
  {\bfseries Event Horizon Telescope} Collaboration, K.~Akiyama {\em et~al.},
    ``{First M87 Event Horizon Telescope Results. I. The Shadow of the
    Supermassive Black Hole},''
    \href{http://dx.doi.org/10.3847/2041-8213/ab0ec7}{{\em Astrophys. J. Lett.}
    {\bfseries 875} (2019) L1}, \href{http://arxiv.org/abs/1906.11238}{{\ttfamily
    arXiv:1906.11238 [astro-ph.GA]}}.
  
  \bibitem{Zeldovich1966}
  Y.~B. Zel'dovich and I.~D. Novikov {\em Astron. Zh.} {\bfseries 43} (1966) 758.
    [Sov. Astron. 10, (1967) 602].
  
  \bibitem{Hawking:1971ei}
  S.~Hawking, ``{Gravitationally collapsed objects of very low mass},'' {\em Mon.
    Not. Roy. Astron. Soc.} {\bfseries 152} (1971) 75.
  
  \bibitem{Carr:1974nx}
  B.~J. Carr and S.~W. Hawking, ``{Black holes in the early Universe},'' {\em
    Mon. Not. Roy. Astron. Soc.} {\bfseries 168} (1974) 399--415.
  
  \bibitem{Carr:2009jm}
  B.~J. Carr, K.~Kohri, Y.~Sendouda, and J.~Yokoyama, ``{New cosmological
    constraints on primordial black holes},''
    \href{http://dx.doi.org/10.1103/PhysRevD.81.104019}{{\em Phys. Rev. D}
    {\bfseries 81} (2010) 104019},
    \href{http://arxiv.org/abs/0912.5297}{{\ttfamily arXiv:0912.5297
    [astro-ph.CO]}}.
  
  \bibitem{Carr:2020gox}
  B.~Carr, K.~Kohri, Y.~Sendouda, and J.~Yokoyama, ``{Constraints on primordial
    black holes},'' \href{http://dx.doi.org/10.1088/1361-6633/ac1e31}{{\em Rept.
    Prog. Phys.} {\bfseries 84} no.~11, (2021) 116902},
    \href{http://arxiv.org/abs/2002.12778}{{\ttfamily arXiv:2002.12778
    [astro-ph.CO]}}.
  
  \bibitem{Carr:2020xqk}
  B.~Carr and F.~Kuhnel, ``{Primordial Black Holes as Dark Matter: Recent
    Developments},''
    \href{http://dx.doi.org/10.1146/annurev-nucl-050520-125911}{{\em Ann. Rev.
    Nucl. Part. Sci.} {\bfseries 70} (2020) 355--394},
    \href{http://arxiv.org/abs/2006.02838}{{\ttfamily arXiv:2006.02838
    [astro-ph.CO]}}.
  
  \bibitem{Escriva:2022duf}
  A.~Escriv\`a, F.~Kuhnel, and Y.~Tada, ``{Primordial Black Holes},''
    \href{http://arxiv.org/abs/2211.05767}{{\ttfamily arXiv:2211.05767
    [astro-ph.CO]}}.
  
  \bibitem{Bird:2016dcv}
  S.~Bird, I.~Cholis, J.~B. Mu\~noz, Y.~Ali-Ha\"\i{}moud, M.~Kamionkowski, E.~D.
    Kovetz, A.~Raccanelli, and A.~G. Riess, ``{Did LIGO detect dark matter?},''
    \href{http://dx.doi.org/10.1103/PhysRevLett.116.201301}{{\em Phys. Rev.
    Lett.} {\bfseries 116} no.~20, (2016) 201301},
    \href{http://arxiv.org/abs/1603.00464}{{\ttfamily arXiv:1603.00464
    [astro-ph.CO]}}.
  
  \bibitem{Clesse:2016vqa}
  S.~Clesse and J.~Garc\'\i{}a-Bellido, ``{The clustering of massive Primordial
    Black Holes as Dark Matter: measuring their mass distribution with Advanced
    LIGO},'' \href{http://dx.doi.org/10.1016/j.dark.2016.10.002}{{\em Phys. Dark
    Univ.} {\bfseries 15} (2017) 142--147},
    \href{http://arxiv.org/abs/1603.05234}{{\ttfamily arXiv:1603.05234
    [astro-ph.CO]}}.
  
  \bibitem{Sasaki:2016jop}
  M.~Sasaki, T.~Suyama, T.~Tanaka, and S.~Yokoyama, ``{Primordial Black Hole
    Scenario for the Gravitational-Wave Event GW150914},''
    \href{http://dx.doi.org/10.1103/PhysRevLett.117.061101}{{\em Phys. Rev.
    Lett.} {\bfseries 117} no.~6, (2016) 061101},
    \href{http://arxiv.org/abs/1603.08338}{{\ttfamily arXiv:1603.08338
    [astro-ph.CO]}}. [Erratum: Phys.Rev.Lett. 121, 059901 (2018)].
  
  \bibitem{Sasaki:2018dmp}
  M.~Sasaki, T.~Suyama, T.~Tanaka, and S.~Yokoyama, ``{Primordial black
    holes\textemdash{}perspectives in gravitational wave astronomy},''
    \href{http://dx.doi.org/10.1088/1361-6382/aaa7b4}{{\em Class. Quant. Grav.}
    {\bfseries 35} no.~6, (2018) 063001},
    \href{http://arxiv.org/abs/1801.05235}{{\ttfamily arXiv:1801.05235
    [astro-ph.CO]}}.
  
  \bibitem{Niikura:2019kqi}
  H.~Niikura, M.~Takada, S.~Yokoyama, T.~Sumi, and S.~Masaki, ``{Constraints on
    Earth-mass primordial black holes from OGLE 5-year microlensing events},''
    \href{http://dx.doi.org/10.1103/PhysRevD.99.083503}{{\em Phys. Rev. D}
    {\bfseries 99} no.~8, (2019) 083503},
    \href{http://arxiv.org/abs/1901.07120}{{\ttfamily arXiv:1901.07120
    [astro-ph.CO]}}.
  
  \bibitem{NANOGrav:2020bcs}
  {\bfseries NANOGrav} Collaboration, Z.~Arzoumanian {\em et~al.}, ``{The
    NANOGrav 12.5 yr Data Set: Search for an Isotropic Stochastic
    Gravitational-wave Background},''
    \href{http://dx.doi.org/10.3847/2041-8213/abd401}{{\em Astrophys. J. Lett.}
    {\bfseries 905} no.~2, (2020) L34},
    \href{http://arxiv.org/abs/2009.04496}{{\ttfamily arXiv:2009.04496
    [astro-ph.HE]}}.
  
  \bibitem{Vaskonen:2020lbd}
  V.~Vaskonen and H.~Veerm\"ae, ``{Did NANOGrav see a signal from primordial
    black hole formation?},''
    \href{http://dx.doi.org/10.1103/PhysRevLett.126.051303}{{\em Phys. Rev.
    Lett.} {\bfseries 126} no.~5, (2021) 051303},
    \href{http://arxiv.org/abs/2009.07832}{{\ttfamily arXiv:2009.07832
    [astro-ph.CO]}}.
  
  \bibitem{DeLuca:2020agl}
  V.~De~Luca, G.~Franciolini, and A.~Riotto, ``{NANOGrav Data Hints at Primordial
    Black Holes as Dark Matter},''
    \href{http://dx.doi.org/10.1103/PhysRevLett.126.041303}{{\em Phys. Rev.
    Lett.} {\bfseries 126} no.~4, (2021) 041303},
    \href{http://arxiv.org/abs/2009.08268}{{\ttfamily arXiv:2009.08268
    [astro-ph.CO]}}.
  
  \bibitem{Kohri:2020qqd}
  K.~Kohri and T.~Terada, ``{Solar-Mass Primordial Black Holes Explain NANOGrav
    Hint of Gravitational Waves},''
    \href{http://dx.doi.org/10.1016/j.physletb.2020.136040}{{\em Phys. Lett. B}
    {\bfseries 813} (2021) 136040},
    \href{http://arxiv.org/abs/2009.11853}{{\ttfamily arXiv:2009.11853
    [astro-ph.CO]}}.
  
  \bibitem{Sugiyama:2020roc}
  S.~Sugiyama, V.~Takhistov, E.~Vitagliano, A.~Kusenko, M.~Sasaki, and M.~Takada,
    ``{Testing Stochastic Gravitational Wave Signals from Primordial Black Holes
    with Optical Telescopes},''
    \href{http://dx.doi.org/10.1016/j.physletb.2021.136097}{{\em Phys. Lett. B}
    {\bfseries 814} (2021) 136097},
    \href{http://arxiv.org/abs/2010.02189}{{\ttfamily arXiv:2010.02189
    [astro-ph.CO]}}.
  
  \bibitem{Domenech:2020ers}
  G.~Dom\`enech and S.~Pi, ``{NANOGrav Hints on Planet-Mass Primordial Black
    Holes},'' \href{http://arxiv.org/abs/2010.03976}{{\ttfamily arXiv:2010.03976
    [astro-ph.CO]}}.
  
  \bibitem{Inomata:2020xad}
  K.~Inomata, M.~Kawasaki, K.~Mukaida, and T.~T. Yanagida, ``{NANOGrav Results
    and LIGO-Virgo Primordial Black Holes in Axionlike Curvaton Models},''
    \href{http://dx.doi.org/10.1103/PhysRevLett.126.131301}{{\em Phys. Rev.
    Lett.} {\bfseries 126} no.~13, (2021) 131301},
    \href{http://arxiv.org/abs/2011.01270}{{\ttfamily arXiv:2011.01270
    [astro-ph.CO]}}.
  
  \bibitem{Kawasaki:2012kn}
  M.~Kawasaki, A.~Kusenko, and T.~T. Yanagida, ``{Primordial seeds of
    supermassive black holes},''
    \href{http://dx.doi.org/10.1016/j.physletb.2012.03.056}{{\em Phys. Lett. B}
    {\bfseries 711} (2012) 1--5},
    \href{http://arxiv.org/abs/1202.3848}{{\ttfamily arXiv:1202.3848
    [astro-ph.CO]}}.
  
  \bibitem{Kohri:2014lza}
  K.~Kohri, T.~Nakama, and T.~Suyama, ``{Testing scenarios of primordial black
    holes being the seeds of supermassive black holes by ultracompact minihalos
    and CMB $\mu$-distortions},''
    \href{http://dx.doi.org/10.1103/PhysRevD.90.083514}{{\em Phys. Rev. D}
    {\bfseries 90} no.~8, (2014) 083514},
    \href{http://arxiv.org/abs/1405.5999}{{\ttfamily arXiv:1405.5999
    [astro-ph.CO]}}.
  
  \bibitem{Nakama:2016kfq}
  T.~Nakama, T.~Suyama, and J.~Yokoyama, ``{Supermassive black holes formed by
    direct collapse of inflationary perturbations},''
    \href{http://dx.doi.org/10.1103/PhysRevD.94.103522}{{\em Phys. Rev. D}
    {\bfseries 94} no.~10, (2016) 103522},
    \href{http://arxiv.org/abs/1609.02245}{{\ttfamily arXiv:1609.02245 [gr-qc]}}.
  
  \bibitem{Carr:2018rid}
  B.~Carr and J.~Silk, ``{Primordial Black Holes as Generators of Cosmic
    Structures},'' \href{http://dx.doi.org/10.1093/mnras/sty1204}{{\em Mon. Not.
    Roy. Astron. Soc.} {\bfseries 478} no.~3, (2018) 3756--3775},
    \href{http://arxiv.org/abs/1801.00672}{{\ttfamily arXiv:1801.00672
    [astro-ph.CO]}}.
  
  \bibitem{Serpico:2020ehh}
  P.~D. Serpico, V.~Poulin, D.~Inman, and K.~Kohri, ``{Cosmic microwave
    background bounds on primordial black holes including dark matter halo
    accretion},'' \href{http://dx.doi.org/10.1103/PhysRevResearch.2.023204}{{\em
    Phys. Rev. Res.} {\bfseries 2} no.~2, (2020) 023204},
    \href{http://arxiv.org/abs/2002.10771}{{\ttfamily arXiv:2002.10771
    [astro-ph.CO]}}.
  
  \bibitem{Unal:2020mts}
  C.~\"Unal, E.~D. Kovetz, and S.~P. Patil, ``{Multimessenger probes of
    inflationary fluctuations and primordial black holes},''
    \href{http://dx.doi.org/10.1103/PhysRevD.103.063519}{{\em Phys. Rev. D}
    {\bfseries 103} no.~6, (2021) 063519},
    \href{http://arxiv.org/abs/2008.11184}{{\ttfamily arXiv:2008.11184
    [astro-ph.CO]}}.
  
  \bibitem{Kohri:2022wzp}
  K.~Kohri, T.~Sekiguchi, and S.~Wang, ``{Cosmological 21-cm line observations to
    test scenarios of super-Eddington accretion on to black holes being seeds of
    high-redshifted supermassive black holes},''
    \href{http://dx.doi.org/10.1103/PhysRevD.106.043539}{{\em Phys. Rev. D}
    {\bfseries 106} no.~4, (2022) 043539},
    \href{http://arxiv.org/abs/2201.05300}{{\ttfamily arXiv:2201.05300
    [astro-ph.CO]}}.
  
  \bibitem{Toussaint:1978br}
  D.~Toussaint, S.~B. Treiman, F.~Wilczek, and A.~Zee, ``{Matter - Antimatter
    Accounting, Thermodynamics, and Black Hole Radiation},''
    \href{http://dx.doi.org/10.1103/PhysRevD.19.1036}{{\em Phys. Rev. D}
    {\bfseries 19} (1979) 1036--1045}.
  
  \bibitem{Barrow:1990he}
  J.~D. Barrow, E.~J. Copeland, E.~W. Kolb, and A.~R. Liddle, ``{Baryogenesis in
    extended inflation. 2. Baryogenesis via primordial black holes},''
    \href{http://dx.doi.org/10.1103/PhysRevD.43.984}{{\em Phys. Rev. D}
    {\bfseries 43} (1991) 984--994}.
  
  \bibitem{Baumann:2007yr}
  D.~Baumann, P.~J. Steinhardt, and N.~Turok, ``{Primordial Black Hole
    Baryogenesis},'' \href{http://arxiv.org/abs/hep-th/0703250}{{\ttfamily
    arXiv:hep-th/0703250}}.
  
  \bibitem{Inomata:2020lmk}
  K.~Inomata, M.~Kawasaki, K.~Mukaida, T.~Terada, and T.~T. Yanagida,
    ``{Gravitational Wave Production right after a Primordial Black Hole
    Evaporation},'' \href{http://dx.doi.org/10.1103/PhysRevD.101.123533}{{\em
    Phys. Rev. D} {\bfseries 101} no.~12, (2020) 123533},
    \href{http://arxiv.org/abs/2003.10455}{{\ttfamily arXiv:2003.10455
    [astro-ph.CO]}}.
  
  \bibitem{Papanikolaou:2020qtd}
  T.~Papanikolaou, V.~Vennin, and D.~Langlois, ``{Gravitational waves from a
    universe filled with primordial black holes},''
    \href{http://dx.doi.org/10.1088/1475-7516/2021/03/053}{{\em JCAP} {\bfseries
    03} (2021) 053}, \href{http://arxiv.org/abs/2010.11573}{{\ttfamily
    arXiv:2010.11573 [astro-ph.CO]}}.
  
  \bibitem{Domenech:2020ssp}
  G.~Dom\`enech, C.~Lin, and M.~Sasaki, ``{Gravitational wave constraints on the
    primordial black hole dominated early universe},''
    \href{http://dx.doi.org/10.1088/1475-7516/2021/04/062}{{\em JCAP} {\bfseries
    04} (2021) 062}, \href{http://arxiv.org/abs/2012.08151}{{\ttfamily
    arXiv:2012.08151 [gr-qc]}}.
  
  \bibitem{Domenech:2021wkk}
  G.~Dom\`enech, V.~Takhistov, and M.~Sasaki, ``{Exploring Evaporating Primordial
    Black Holes with Gravitational Waves},''
    \href{http://arxiv.org/abs/2105.06816}{{\ttfamily arXiv:2105.06816
    [astro-ph.CO]}}.
  
  \bibitem{Bhaumik:2020dor}
  N.~Bhaumik and R.~K. Jain, ``{Small scale induced gravitational waves from
    primordial black holes, a~stringent lower mass bound, and the imprints of an
    early matter to~radiation transition},''
    \href{http://dx.doi.org/10.1103/PhysRevD.104.023531}{{\em Phys. Rev. D}
    {\bfseries 104} no.~2, (2021) 023531},
    \href{http://arxiv.org/abs/2009.10424}{{\ttfamily arXiv:2009.10424
    [astro-ph.CO]}}.
  
  \bibitem{Hooper:2020evu}
  D.~Hooper, G.~Krnjaic, J.~March-Russell, S.~D. McDermott, and
    R.~Petrossian-Byrne, ``{Hot Gravitons and Gravitational Waves From Kerr Black
    Holes in the Early Universe},''
    \href{http://arxiv.org/abs/2004.00618}{{\ttfamily arXiv:2004.00618
    [astro-ph.CO]}}.
  
  \bibitem{Inomata:2019ivs}
  K.~Inomata, K.~Kohri, T.~Nakama, and T.~Terada, ``{Enhancement of Gravitational
    Waves Induced by Scalar Perturbations due to a Sudden Transition from an
    Early Matter Era to the Radiation Era},''
    \href{http://dx.doi.org/10.1103/PhysRevD.100.043532}{{\em Phys. Rev. D}
    {\bfseries 100} no.~4, (2019) 043532},
    \href{http://arxiv.org/abs/1904.12879}{{\ttfamily arXiv:1904.12879
    [astro-ph.CO]}}.
  
  \bibitem{Papanikolaou:2022chm}
  T.~Papanikolaou, ``{Gravitational waves induced from primordial black hole
    fluctuations: the~effect of an extended mass function},''
    \href{http://dx.doi.org/10.1088/1475-7516/2022/10/089}{{\em JCAP} {\bfseries
    10} (2022) 089}, \href{http://arxiv.org/abs/2207.11041}{{\ttfamily
    arXiv:2207.11041 [astro-ph.CO]}}.
  
  \bibitem{Ananda:2006af}
  K.~N. Ananda, C.~Clarkson, and D.~Wands, ``{The Cosmological gravitational wave
    background from primordial density perturbations},''
    \href{http://dx.doi.org/10.1103/PhysRevD.75.123518}{{\em Phys. Rev. D}
    {\bfseries 75} (2007) 123518},
    \href{http://arxiv.org/abs/gr-qc/0612013}{{\ttfamily arXiv:gr-qc/0612013}}.
  
  \bibitem{Baumann:2007zm}
  D.~Baumann, P.~J. Steinhardt, K.~Takahashi, and K.~Ichiki, ``{Gravitational
    Wave Spectrum Induced by Primordial Scalar Perturbations},''
    \href{http://dx.doi.org/10.1103/PhysRevD.76.084019}{{\em Phys. Rev. D}
    {\bfseries 76} (2007) 084019},
    \href{http://arxiv.org/abs/hep-th/0703290}{{\ttfamily arXiv:hep-th/0703290}}.
  
  \bibitem{Saito:2008jc}
  R.~Saito and J.~Yokoyama, ``{Gravitational wave background as a probe of the
    primordial black hole abundance},''
    \href{http://dx.doi.org/10.1103/PhysRevLett.102.161101}{{\em Phys. Rev.
    Lett.} {\bfseries 102} (2009) 161101},
    \href{http://arxiv.org/abs/0812.4339}{{\ttfamily arXiv:0812.4339
    [astro-ph]}}. [Erratum: Phys.Rev.Lett. 107, 069901 (2011)].
  
  \bibitem{Saito:2009jt}
  R.~Saito and J.~Yokoyama, ``{Gravitational-Wave Constraints on the Abundance of
    Primordial Black Holes},'' \href{http://dx.doi.org/10.1143/PTP.126.351}{{\em
    Prog. Theor. Phys.} {\bfseries 123} (2010) 867--886},
    \href{http://arxiv.org/abs/0912.5317}{{\ttfamily arXiv:0912.5317
    [astro-ph.CO]}}. [Erratum: Prog.Theor.Phys. 126, 351--352 (2011)].
  
  \bibitem{Assadullahi:2009nf}
  H.~Assadullahi and D.~Wands, ``{Gravitational waves from an early matter
    era},'' \href{http://dx.doi.org/10.1103/PhysRevD.79.083511}{{\em Phys. Rev.
    D} {\bfseries 79} (2009) 083511},
    \href{http://arxiv.org/abs/0901.0989}{{\ttfamily arXiv:0901.0989
    [astro-ph.CO]}}.
  
  \bibitem{Espinosa:2018eve}
  J.~R. Espinosa, D.~Racco, and A.~Riotto, ``{A Cosmological Signature of the SM
    Higgs Instability: Gravitational Waves},''
    \href{http://dx.doi.org/10.1088/1475-7516/2018/09/012}{{\em JCAP} {\bfseries
    09} (2018) 012}, \href{http://arxiv.org/abs/1804.07732}{{\ttfamily
    arXiv:1804.07732 [hep-ph]}}.
  
  \bibitem{Kohri:2018awv}
  K.~Kohri and T.~Terada, ``{Semianalytic calculation of gravitational wave
    spectrum nonlinearly induced from primordial curvature perturbations},''
    \href{http://dx.doi.org/10.1103/PhysRevD.97.123532}{{\em Phys. Rev. D}
    {\bfseries 97} no.~12, (2018) 123532},
    \href{http://arxiv.org/abs/1804.08577}{{\ttfamily arXiv:1804.08577 [gr-qc]}}.
  
  \bibitem{Cai:2018dig}
  R.-g. Cai, S.~Pi, and M.~Sasaki, ``{Gravitational Waves Induced by non-Gaussian
    Scalar Perturbations},''
    \href{http://dx.doi.org/10.1103/PhysRevLett.122.201101}{{\em Phys. Rev.
    Lett.} {\bfseries 122} no.~20, (2019) 201101},
    \href{http://arxiv.org/abs/1810.11000}{{\ttfamily arXiv:1810.11000
    [astro-ph.CO]}}.
  
  \bibitem{Yuan:2021qgz}
  C.~Yuan and Q.-G. Huang, ``{A topic review on probing primordial black hole
    dark matter with scalar induced gravitational waves},''
    \href{http://arxiv.org/abs/2103.04739}{{\ttfamily arXiv:2103.04739
    [astro-ph.GA]}}.
  
  \bibitem{Domenech:2021ztg}
  G.~Dom\`enech, ``{Scalar induced gravitational waves review},''
    \href{http://arxiv.org/abs/2109.01398}{{\ttfamily arXiv:2109.01398 [gr-qc]}}.
  
  \bibitem{Shibata:1999zs}
  M.~Shibata and M.~Sasaki, ``{Black hole formation in the Friedmann universe:
    Formulation and computation in numerical relativity},''
    \href{http://dx.doi.org/10.1103/PhysRevD.60.084002}{{\em Phys. Rev. D}
    {\bfseries 60} (1999) 084002},
    \href{http://arxiv.org/abs/gr-qc/9905064}{{\ttfamily arXiv:gr-qc/9905064}}.
  
  \bibitem{Musco:2004ak}
  I.~Musco, J.~C. Miller, and L.~Rezzolla, ``{Computations of primordial black
    hole formation},'' \href{http://dx.doi.org/10.1088/0264-9381/22/7/013}{{\em
    Class. Quant. Grav.} {\bfseries 22} (2005) 1405--1424},
    \href{http://arxiv.org/abs/gr-qc/0412063}{{\ttfamily arXiv:gr-qc/0412063}}.
  
  \bibitem{Polnarev:2006aa}
  A.~G. Polnarev and I.~Musco, ``{Curvature profiles as initial conditions for
    primordial black hole formation},''
    \href{http://dx.doi.org/10.1088/0264-9381/24/6/003}{{\em Class. Quant. Grav.}
    {\bfseries 24} (2007) 1405--1432},
    \href{http://arxiv.org/abs/gr-qc/0605122}{{\ttfamily arXiv:gr-qc/0605122}}.
  
  \bibitem{Musco:2008hv}
  I.~Musco, J.~C. Miller, and A.~G. Polnarev, ``{Primordial black hole formation
    in the radiative era: Investigation of the critical nature of the
    collapse},'' \href{http://dx.doi.org/10.1088/0264-9381/26/23/235001}{{\em
    Class. Quant. Grav.} {\bfseries 26} (2009) 235001},
    \href{http://arxiv.org/abs/0811.1452}{{\ttfamily arXiv:0811.1452 [gr-qc]}}.
  
  \bibitem{Musco:2012au}
  I.~Musco and J.~C. Miller, ``{Primordial black hole formation in the early
    universe: critical behaviour and self-similarity},''
    \href{http://dx.doi.org/10.1088/0264-9381/30/14/145009}{{\em Class. Quant.
    Grav.} {\bfseries 30} (2013) 145009},
    \href{http://arxiv.org/abs/1201.2379}{{\ttfamily arXiv:1201.2379 [gr-qc]}}.
  
  \bibitem{Nakama:2013ica}
  T.~Nakama, T.~Harada, A.~G. Polnarev, and J.~Yokoyama, ``{Identifying the most
    crucial parameters of the initial curvature profile for primordial black hole
    formation},'' \href{http://dx.doi.org/10.1088/1475-7516/2014/01/037}{{\em
    JCAP} {\bfseries 01} (2014) 037},
    \href{http://arxiv.org/abs/1310.3007}{{\ttfamily arXiv:1310.3007 [gr-qc]}}.
  
  \bibitem{Harada:2013epa}
  T.~Harada, C.-M. Yoo, and K.~Kohri, ``{Threshold of primordial black hole
    formation},'' \href{http://dx.doi.org/10.1103/PhysRevD.88.084051}{{\em Phys.
    Rev. D} {\bfseries 88} no.~8, (2013) 084051},
    \href{http://arxiv.org/abs/1309.4201}{{\ttfamily arXiv:1309.4201
    [astro-ph.CO]}}. [Erratum: Phys.Rev.D 89, 029903 (2014)].
  
  \bibitem{Harada:2015yda}
  T.~Harada, C.-M. Yoo, T.~Nakama, and Y.~Koga, ``{Cosmological long-wavelength
    solutions and primordial black hole formation},''
    \href{http://dx.doi.org/10.1103/PhysRevD.91.084057}{{\em Phys. Rev. D}
    {\bfseries 91} no.~8, (2015) 084057},
    \href{http://arxiv.org/abs/1503.03934}{{\ttfamily arXiv:1503.03934 [gr-qc]}}.
  
  \bibitem{Musco:2018rwt}
  I.~Musco, ``{Threshold for primordial black holes: Dependence on the shape of
    the cosmological perturbations},''
    \href{http://dx.doi.org/10.1103/PhysRevD.100.123524}{{\em Phys. Rev. D}
    {\bfseries 100} no.~12, (2019) 123524},
    \href{http://arxiv.org/abs/1809.02127}{{\ttfamily arXiv:1809.02127 [gr-qc]}}.
  
  \bibitem{Escriva:2019phb}
  A.~Escriv\`a, C.~Germani, and R.~K. Sheth, ``{Universal threshold for
    primordial black hole formation},''
    \href{http://dx.doi.org/10.1103/PhysRevD.101.044022}{{\em Phys. Rev. D}
    {\bfseries 101} no.~4, (2020) 044022},
    \href{http://arxiv.org/abs/1907.13311}{{\ttfamily arXiv:1907.13311 [gr-qc]}}.
  
  \bibitem{Musco:2020jjb}
  I.~Musco, V.~De~Luca, G.~Franciolini, and A.~Riotto, ``{Threshold for
    primordial black holes. II. A simple analytic prescription},''
    \href{http://dx.doi.org/10.1103/PhysRevD.103.063538}{{\em Phys. Rev. D}
    {\bfseries 103} no.~6, (2021) 063538},
    \href{http://arxiv.org/abs/2011.03014}{{\ttfamily arXiv:2011.03014
    [astro-ph.CO]}}.
  
  \bibitem{Musco:2021sva}
  I.~Musco and T.~Papanikolaou, ``{Primordial black hole formation for an
    anisotropic perfect fluid: Initial conditions and estimation of the
    threshold},'' \href{http://dx.doi.org/10.1103/PhysRevD.106.083017}{{\em Phys.
    Rev. D} {\bfseries 106} no.~8, (2022) 083017},
    \href{http://arxiv.org/abs/2110.05982}{{\ttfamily arXiv:2110.05982 [gr-qc]}}.
  
  \bibitem{Papanikolaou:2022cvo}
  T.~Papanikolaou, ``{Toward the primordial black hole formation threshold in a
    time-dependent equation-of-state background},''
    \href{http://dx.doi.org/10.1103/PhysRevD.105.124055}{{\em Phys. Rev. D}
    {\bfseries 105} no.~12, (2022) 124055},
    \href{http://arxiv.org/abs/2205.07748}{{\ttfamily arXiv:2205.07748 [gr-qc]}}.
  
  \bibitem{1986ApJ...304...15B}
  J.~M. {Bardeen}, J.~R. {Bond}, N.~{Kaiser}, and A.~S. {Szalay}, ``{The
    statistics of peaks of Gaussian random fields},''
    \href{http://dx.doi.org/10.1086/164143}{{\em Astrophys. J.} {\bfseries 304}
    (May, 1986) 15--61}.
  
  \bibitem{Yoo:2018kvb}
  C.-M. Yoo, T.~Harada, J.~Garriga, and K.~Kohri, ``{Primordial black hole
    abundance from random Gaussian curvature perturbations and a local density
    threshold},'' \href{http://dx.doi.org/10.1093/ptep/pty120}{{\em PTEP}
    {\bfseries 2018} no.~12, (2018) 123E01},
    \href{http://arxiv.org/abs/1805.03946}{{\ttfamily arXiv:1805.03946
    [astro-ph.CO]}}.
  
  \bibitem{Germani:2018jgr}
  C.~Germani and I.~Musco, ``{Abundance of Primordial Black Holes Depends on the
    Shape of the Inflationary Power Spectrum},''
    \href{http://dx.doi.org/10.1103/PhysRevLett.122.141302}{{\em Phys. Rev.
    Lett.} {\bfseries 122} no.~14, (2019) 141302},
    \href{http://arxiv.org/abs/1805.04087}{{\ttfamily arXiv:1805.04087
    [astro-ph.CO]}}.
  
  \bibitem{Yoo:2020dkz}
  C.-M. Yoo, T.~Harada, S.~Hirano, and K.~Kohri, ``{Abundance of Primordial Black
    Holes in Peak Theory for an Arbitrary Power Spectrum},''
    \href{http://dx.doi.org/10.1093/ptep/ptaa155}{{\em PTEP} {\bfseries 2021}
    no.~1, (2021) 013E02}, \href{http://arxiv.org/abs/2008.02425}{{\ttfamily
    arXiv:2008.02425 [astro-ph.CO]}}.
  
  \bibitem{Niemeyer:1997mt}
  J.~C. Niemeyer and K.~Jedamzik, ``{Near-critical gravitational collapse and the
    initial mass function of primordial black holes},''
    \href{http://dx.doi.org/10.1103/PhysRevLett.80.5481}{{\em Phys. Rev. Lett.}
    {\bfseries 80} (1998) 5481--5484},
    \href{http://arxiv.org/abs/astro-ph/9709072}{{\ttfamily
    arXiv:astro-ph/9709072}}.
  
  \bibitem{Yokoyama:1998xd}
  J.~Yokoyama, ``{Cosmological constraints on primordial black holes produced in
    the near critical gravitational collapse},''
    \href{http://dx.doi.org/10.1103/PhysRevD.58.107502}{{\em Phys. Rev. D}
    {\bfseries 58} (1998) 107502},
    \href{http://arxiv.org/abs/gr-qc/9804041}{{\ttfamily arXiv:gr-qc/9804041}}.
  
  \bibitem{Green:1999xm}
  A.~M. Green and A.~R. Liddle, ``{Critical collapse and the primordial black
    hole initial mass function},''
    \href{http://dx.doi.org/10.1103/PhysRevD.60.063509}{{\em Phys. Rev. D}
    {\bfseries 60} (1999) 063509},
    \href{http://arxiv.org/abs/astro-ph/9901268}{{\ttfamily
    arXiv:astro-ph/9901268}}.
  
  \bibitem{Kuhnel:2015vtw}
  F.~K\"uhnel, C.~Rampf, and M.~Sandstad, ``{Effects of Critical Collapse on
    Primordial Black-Hole Mass Spectra},''
    \href{http://dx.doi.org/10.1140/epjc/s10052-016-3945-8}{{\em Eur. Phys. J. C}
    {\bfseries 76} no.~2, (2016) 93},
    \href{http://arxiv.org/abs/1512.00488}{{\ttfamily arXiv:1512.00488
    [astro-ph.CO]}}.
  
  \bibitem{Kitajima:2021fpq}
  N.~Kitajima, Y.~Tada, S.~Yokoyama, and C.-M. Yoo, ``{Primordial black holes in
    peak theory with a non-Gaussian tail},''
    \href{http://dx.doi.org/10.1088/1475-7516/2021/10/053}{{\em JCAP} {\bfseries
    10} (2021) 053}, \href{http://arxiv.org/abs/2109.00791}{{\ttfamily
    arXiv:2109.00791 [astro-ph.CO]}}.
  
  \bibitem{Atal:2019erb}
  V.~Atal, J.~Cid, A.~Escriv\`a, and J.~Garriga, ``{PBH in single field
    inflation: the effect of shape dispersion and non-Gaussianities},''
    \href{http://dx.doi.org/10.1088/1475-7516/2020/05/022}{{\em JCAP} {\bfseries
    05} (2020) 022}, \href{http://arxiv.org/abs/1908.11357}{{\ttfamily
    arXiv:1908.11357 [astro-ph.CO]}}.
  
  \bibitem{Yoo:2019pma}
  C.-M. Yoo, J.-O. Gong, and S.~Yokoyama, ``{Abundance of primordial black holes
    with local non-Gaussianity in peak theory},''
    \href{http://dx.doi.org/10.1088/1475-7516/2019/09/033}{{\em JCAP} {\bfseries
    09} (2019) 033}, \href{http://arxiv.org/abs/1906.06790}{{\ttfamily
    arXiv:1906.06790 [astro-ph.CO]}}.
  
  \bibitem{Atal:2019cdz}
  V.~Atal, J.~Garriga, and A.~Marcos-Caballero, ``{Primordial black hole
    formation with non-Gaussian curvature perturbations},''
    \href{http://dx.doi.org/10.1088/1475-7516/2019/09/073}{{\em JCAP} {\bfseries
    09} (2019) 073}, \href{http://arxiv.org/abs/1905.13202}{{\ttfamily
    arXiv:1905.13202 [astro-ph.CO]}}.
  
  \bibitem{Escriva:2022pnz}
  A.~Escriv\`a, Y.~Tada, S.~Yokoyama, and C.-M. Yoo, ``{Simulation of primordial
    black holes with large negative non-Gaussianity},''
    \href{http://dx.doi.org/10.1088/1475-7516/2022/05/012}{{\em JCAP} {\bfseries
    05} no.~05, (2022) 012}, \href{http://arxiv.org/abs/2202.01028}{{\ttfamily
    arXiv:2202.01028 [astro-ph.CO]}}.
  
  \bibitem{Khlopov:1980mg}
  M.~Y. Khlopov and A.~G. Polnarev, ``{PRIMORDIAL BLACK HOLES AS A COSMOLOGICAL
    TEST OF GRAND UNIFICATION},''
    \href{http://dx.doi.org/10.1016/0370-2693(80)90624-3}{{\em Phys. Lett. B}
    {\bfseries 97} (1980) 383--387}.
  
  \bibitem{Polnarev:1986}
  A.~G. Polnarev and M.~{\relax Yu}. Khlopov, ``{COSMOLOGY, PRIMORDIAL BLACK
    HOLES, AND SUPERMASSIVE PARTICLES},''
    \href{http://dx.doi.org/10.1070/PU1985v028n03ABEH003858}{{\em Sov. Phys.
    Usp.} {\bfseries 28} (1985) 213--232}.
  [Usp. Fiz. Nauk145,369(1985)].
  
  \bibitem{1965ApJ...142.1431L}
  C.~C. {Lin}, L.~{Mestel}, and F.~H. {Shu}, ``{The Gravitational Collapse of a
    Uniform Spheroid.},'' \href{http://dx.doi.org/10.1086/148428}{{\em Astrophys.
    J.} {\bfseries 142} (Nov., 1965) 1431}.
  
  \bibitem{1970A&A.....5...84Z}
  Y.~B. {Zel'Dovich}, ``{Reprint of 1970A\&A.....5...84Z. Gravitational
    instability: an approximate theory for large density perturbations.},'' {\em
    Astron. Astrophys.} {\bfseries 500} (Mar., 1970) 13--18.
  
  \bibitem{Harada:2016mhb}
  T.~Harada, C.-M. Yoo, K.~Kohri, K.-i. Nakao, and S.~Jhingan, ``{Primordial
    black hole formation in the matter-dominated phase of the Universe},''
    \href{http://dx.doi.org/10.3847/1538-4357/833/1/61}{{\em Astrophys. J.}
    {\bfseries 833} no.~1, (2016) 61},
    \href{http://arxiv.org/abs/1609.01588}{{\ttfamily arXiv:1609.01588
    [astro-ph.CO]}}.
  
  \bibitem{Harada:2017fjm}
  T.~Harada, C.-M. Yoo, K.~Kohri, and K.-I. Nakao, ``{Spins of primordial black
    holes formed in the matter-dominated phase of the Universe},''
    \href{http://dx.doi.org/10.1103/PhysRevD.96.083517}{{\em Phys. Rev. D}
    {\bfseries 96} no.~8, (2017) 083517},
    \href{http://arxiv.org/abs/1707.03595}{{\ttfamily arXiv:1707.03595 [gr-qc]}}.
    [Erratum: Phys.Rev.D 99, 069904 (2019)].
  
  \bibitem{Kokubu:2018fxy}
  T.~Kokubu, K.~Kyutoku, K.~Kohri, and T.~Harada, ``{Effect of Inhomogeneity on
    Primordial Black Hole Formation in the Matter Dominated Era},''
    \href{http://dx.doi.org/10.1103/PhysRevD.98.123024}{{\em Phys. Rev. D}
    {\bfseries 98} no.~12, (2018) 123024},
    \href{http://arxiv.org/abs/1810.03490}{{\ttfamily arXiv:1810.03490
    [astro-ph.CO]}}.
  
  \bibitem{deJong:2021bbo}
  E.~de~Jong, J.~C. Aurrekoetxea, and E.~A. Lim, ``{Primordial black hole
    formation with full numerical relativity},''
    \href{http://arxiv.org/abs/2109.04896}{{\ttfamily arXiv:2109.04896
    [astro-ph.CO]}}.
  
  \bibitem{Padilla:2021zgm}
  L.~E. Padilla, J.~C. Hidalgo, and K.~A. Malik, ``{A new mechanism for
    primordial black hole formation during reheating},''
    \href{http://arxiv.org/abs/2110.14584}{{\ttfamily arXiv:2110.14584
    [astro-ph.CO]}}.
  
  \bibitem{Yoo:2022mzl}
  C.-M. Yoo, ``{The basics of primordial black hole formation and abundance
    estimation},'' \href{http://dx.doi.org/10.3390/galaxies10060112}{{\em
    Galaxies} {\bfseries 10} no.~6, (2022) },
    \href{http://arxiv.org/abs/2211.13512}{{\ttfamily arXiv:2211.13512
    [astro-ph.CO]}}.
  
  \bibitem{Klauder:1972lsv}
  K.~Thorne, {\em {Nonspherical Gravitational Collapse: a Short Review {\rm in:
    }Magic without magic: John Archibald Wheeler}: {A collection of essays in
    honor of his sixtieth birthday}}.
  \newblock Freeman, San Francisco, 1972.
  
  \bibitem{Doroshkevich:1970}
  A.~G. Doroshkevich, ``{Spatial structure of perturbations and origin of
    galactic rotation in fluctuation theory},''
    \href{http://dx.doi.org/10.1007/BF01001625}{{\em Astrophysics} {\bfseries 6}
    (1970) 320}.
  
  \bibitem{Carr:1975qj}
  B.~J. Carr, ``{The Primordial black hole mass spectrum},''
    \href{http://dx.doi.org/10.1086/153853}{{\em Astrophys. J.} {\bfseries 201}
    (1975) 1--19}.
  
  \bibitem{Cotner:2018vug}
  E.~Cotner, A.~Kusenko, and V.~Takhistov, ``{Primordial Black Holes from
    Inflaton Fragmentation into Oscillons},''
    \href{http://dx.doi.org/10.1103/PhysRevD.98.083513}{{\em Phys. Rev. D}
    {\bfseries 98} no.~8, (2018) 083513},
    \href{http://arxiv.org/abs/1801.03321}{{\ttfamily arXiv:1801.03321
    [astro-ph.CO]}}.
  
  \bibitem{Kohri:2018qtx}
  K.~Kohri and T.~Terada, ``{Primordial Black Hole Dark Matter and LIGO/Virgo
    Merger Rate from Inflation with Running Spectral Indices: Formation in the
    Matter- and/or Radiation-Dominated Universe},''
    \href{http://dx.doi.org/10.1088/1361-6382/aaea18}{{\em Class. Quant. Grav.}
    {\bfseries 35} no.~23, (2018) 235017},
    \href{http://arxiv.org/abs/1802.06785}{{\ttfamily arXiv:1802.06785
    [astro-ph.CO]}}.
  
  \bibitem{Martin:2019nuw}
  J.~Martin, T.~Papanikolaou, and V.~Vennin, ``{Primordial black holes from the
    preheating instability in single-field inflation},''
    \href{http://dx.doi.org/10.1088/1475-7516/2020/01/024}{{\em JCAP} {\bfseries
    01} (2020) 024}, \href{http://arxiv.org/abs/1907.04236}{{\ttfamily
    arXiv:1907.04236 [astro-ph.CO]}}.
  
  \bibitem{Martin:2020fgl}
  J.~Martin, T.~Papanikolaou, L.~Pinol, and V.~Vennin, ``{Metric preheating and
    radiative decay in single-field inflation},''
    \href{http://dx.doi.org/10.1088/1475-7516/2020/05/003}{{\em JCAP} {\bfseries
    05} (2020) 003}, \href{http://arxiv.org/abs/2002.01820}{{\ttfamily
    arXiv:2002.01820 [astro-ph.CO]}}.
  
  \bibitem{Auclair:2020csm}
  P.~Auclair and V.~Vennin, ``{Primordial black holes from metric preheating:
    mass fraction in the excursion-set approach},''
    \href{http://dx.doi.org/10.1088/1475-7516/2021/02/038}{{\em JCAP} {\bfseries
    02} (2021) 038}, \href{http://arxiv.org/abs/2011.05633}{{\ttfamily
    arXiv:2011.05633 [astro-ph.CO]}}.
  
  \bibitem{Eggemeier:2021smj}
  B.~Eggemeier, B.~Schwabe, J.~C. Niemeyer, and R.~Easther, ``{Gravitational
    Collapse in the Post-Inflationary Universe},''
    \href{http://arxiv.org/abs/2110.15109}{{\ttfamily arXiv:2110.15109
    [astro-ph.CO]}}.
  
  \bibitem{Hidalgo:2022yed}
  J.~C. Hidalgo, L.~E. Padilla, and G.~German, ``{Production of PBHs from
    inflaton structure},'' \href{http://arxiv.org/abs/2208.09462}{{\ttfamily
    arXiv:2208.09462 [astro-ph.CO]}}.
  
  \bibitem{Weinberg:1972kfs}
  S.~Weinberg, {\em {Gravitation and Cosmology}: {Principles and Applications of
    the General Theory of Relativity}}.
  \newblock John Wiley and Sons, New York, 1972.
  
  \end{thebibliography}

\end{document}